\documentclass[10pt,floatfix,
 twocolumn, 
 amsmath,amssymb,
 aps, physrev,
]{revtex4-2}

\usepackage{makecell} % Para forzar saltos de línea en las celdas
\usepackage{graphicx}% Include figure files
\usepackage{dcolumn}% Align table columns on decimal point
\usepackage{bm}% bold math
\usepackage{caption} %Para colocar el caption de las imágenes
\usepackage{mathtools} % Para usar herramientas mathematics
\usepackage{tensor} % Para usar herramientas tensoriales
\usepackage[hidelinks]{hyperref} % Para crear enlaces
\usepackage{amssymb}%Símbolos mathematics
\usepackage{mathtools}%Más simbolos
\usepackage{mathrsfs}%Más símbolos
\usepackage{amsmath} % Para usar la matriz y los símbolos de derivadas parciales
\usepackage{xcolor}%Paquete para el color
\usepackage{array} %Para las tablas
\usepackage{booktabs}%Otra para tablas
\usepackage{subcaption}%Subreferencias
\usepackage{enumitem}

\begin{document}

\title{Einstein-Maxwell-Dilaton Wormholes that meet the Energy Conditions}

\author{Leonel Bixano}
    \email{Contact author: leonel.delacruz@cinvestav.mx}
\author{Tonatiuh Matos}%
 \email{Contact author: tonatiuh.matos@cinvestav.mx}
\affiliation{Departamento de F\'{\i}sica, Centro de Investigaci\'on y de Estudios Avanzados del IPN, Av. I.P.N. 2508, San Pedro Zacatenco, M\'exico 07360, CDMX.
}%

%\affiliation{Departamento de F\'{\i}sica, Centro de Investigaci\'on y de Estudios Avanzados del IPN, Av. I.P.N. 2508, San Pedro Zacatenco, M\'exico 07360, CDMX.}
\date{\today}

\begin{abstract}
    One of the latest predictions of Einstein's theory is the existence of Wormholes (WH).
In this work, we present exact solutions of the Einstein-Maxwell-Dilaton equations representing traversable Wormholes. These solutions satisfy the energy conditions and have a ring singularity satisfying the cosmic censorship of WHs, i.e. we show that, as in previous solutions, geodesics cannot touch the singularity.
We find that the most optimal input regions for the first class of solutions traversing these wormholes are near the poles and near the equatorial plane for the second class. We also find that the solution associated with the first class is physically feasible, while for the second class it presents the problem of not being asymptotically flat when considering a dilatonic-type scalar field. Finally, we give examples of realistic astrophysical objects that could fulfill these conditions.

\end{abstract}

\maketitle

\section{Introduction}
Einstein's theory of gravitation, called general relativity, has predicted a huge set of phenomena that have been observed in the universe. Phenomena such as black holes and gravitational waves, once thought to be impossible to detect, are now detected and even observed.
One of the latest predictions of Einstein's equations that has not been observed is wormholes (WH). They were predicted by Einstein himself but have so far been treated as the stuff of science fiction. 
In \cite{Morris:1988cz} it was shown that if we want a static solution of Einstein's equations without singularities, the source of this solution must violate the Null Energy Conditions (NEC). However, in 1998 it was discovered that the universe is currently in an accelerated phase of expansion, implying that there are some phenomena in the universe that necessarily violate the NEC, giving rise to the possibility that the energy conditions are not as fundamental as we thought. Since this discovery, interest in WHs has increased enormously.
However, in \cite{Gonzalez:2008wd} it was shown that all static and spherically symmetric WHs are unstable, implying that static WHs may not exist. In the paper \cite{Matos:2005uh} it was conjectured that rotating WHs may be stable, arguing that dynamical stability is more feasible than static stability. Later in this paper \cite{Matos:2010pcd} we give a new formulation for finding new exact solutions of the Einstein-Maxwell-Dilaton (EMD) system and find the first families of exact rotational solutions \cite{Matos:2000za}. This work is, after all, a recipe for finding new exact stationary and axially symmetric solutions of the EMD system. Some of these families contain a ring singularity which makes these solutions potentially unphysical, because they violate the Cosmic Censorship Conjecture (CCC). 
Nevertheless, in \cite{Matos:2005uh} we find that these singularities cannot be touched by a geodesic, given rise to a new form of CCC that we called Wormhole Cosmic Censorship Conjecture (WHCCC). We find this result using numerical simulations \cite{Matos:2012gj} and later we showed that using analytic calculations \cite{DelAguila:2021awa}. In other words, these exact solutions have a ring singularity that is not in causal contact with the rest of the universe, it is a new form of CCC.

In the present work we start with a space-time that contains two Killing-vectors and find axial symmetric, stationary exact solutions of the EMD system representing WH and show that these solutions also satisfy the Energy Conditions (EC) and the WHCCC. To do so, we start from the following Lagrangian
\begin{equation}\label{LagrangianoTesis}
    \mathfrak{L}=\sqrt{-g}\left(-R +2\epsilon_0 (\nabla \phi)^2 +e^{-2 \alpha_0 \phi } F^2 \right),
\end{equation}
where $\phi$ is the scalar field, $\alpha_0$ determines the theory we are using. In other words, we have written the field equations in such a way that we can recover different gravitational theories using the parameter $\alpha_0$, we have
\begin{equation}\label{Alpha0}
    \alpha_0^2 = \begin{cases} 
    3 & \Rightarrow \mathfrak{L} \text{ corresponds to}  \\ & \text{Kaluza-Klein lagrangian}  \\
    1 & \Rightarrow \mathfrak{L} \text{ corresponds to Low-Energy}  \\ & \text{Super Strings lagrangian }\\ 
    0 & \Rightarrow \mathfrak{L} \text{ corresponds to}  \\ & \text{Eintein-Maxwell lagrangian}
    \end{cases}
\end{equation}
 and $\epsilon_0$ is a scalar field constant which can take the values $1$ for the dilatonic field and $-1$ for the phantom field. $F_{\mu \nu}$ is the Faraday tensor and $R$ is the Ricci scalar.

Using variational methods, we obtain the following field equations.
\begin{subequations}\label{EcuacionesDeCampoOriginales}
\begin{equation}\label{Eq:Campo1}
    \nabla_\mu \left( e^{-2\alpha_0 \phi} F^{\mu \nu} \right)=0,
\end{equation}
\begin{equation}\label{Eq:Campo2}
    \epsilon_0 \nabla^2 \phi+\frac{\alpha_0}{2}\left( e^{-2\alpha_0 \phi} F^{2} \right)=0,
\end{equation}
\begin{multline}\label{Eq:Campo3}
    R_{\mu \nu}=2\epsilon_0 \nabla_\mu \phi \nabla_\nu \phi \\ + 2 e^{-2\alpha_0 \phi} \left( F_{\mu \sigma} \tensor{F}{_\nu}{^\sigma} -\frac{1}{4} g_{\mu \nu } F^2 \right) .
\end{multline}

\end{subequations}

We begin by presenting the symmetries considered and their implications for the metric and the electromagnetic potential. Once the metric and electromagnetic potential are proposed, the tools to simplify and solve the field equations (\ref{EcuacionesDeCampoOriginales}) will be introduced. The two classes of solutions\footnote{Taken from the article \cite{Matos:2010pcd}.} will be utilized to obtain the metric functions and the electromagnetic four-potential. After obtaining the solutions, the verification equation will be derived that ensures compliance with the field equations (\ref{EcuacionesDeCampoOriginales}) are derived, determining the value of the constant $k_0$, which dictates the choice of the scalar field. All results are summarized in Table [\ref{TablaValoresk}].

In Section 3, two important invariants, the Ricci scalar and the Kretschmann scalar, will be analyzed to determine the behavior of the spacetime generated by the proposed solutions.

Section 4 explores the null energy condition and presents the tools used in the analysis. Subsequently, Section 5 examines the geometry of the wormhole by embedding the hypersurface defined by constant $y=y_0$ and $t=t_0$ in a cylindrical space. The geometry for both solutions will be visualized through graphical representations.

In Section 6, tidal forces are analyzed using the reference frame of the astronaut, determining optimal entry angles into the wormhole to avoid destruction. The behavior of tidal forces will also be graphed for a better understanding.

Section 7 focuses on the tools used to plot geodesics in spheroidal coordinates. Finally, in Section 8, the consequences of adopting SI units in the theory are discussed. This includes implications for the electromagnetic field, angular velocity, throat size of the wormhole, angular momentum per unit of mass, and electric and magnetic charges. Relationships between these quantities will be established to characterize the wormholes. Some examples of wormholes with a mass equivalent to that of the Sun, a pulsar and a supermassive black hole are presented.

\section{Two Classes of Solutions}

We start from a space-time with two Killing vectors $\partial _\varphi,\partial_t$ that represent stationarity and axial symmetry. In such space-times the compatible 4-potential is given by 
\begin{equation}\label{4Potencial}
    A_{\mu}=\bigg[ A_t(x,y),0,0,A_\varphi (x,y) \bigg] .
\end{equation}
We will adopt the stationary metric on the oblate spheroidal coordinates $\big(x,y)$ composed of metric functions $f=f(x,y)$, $\omega=\omega(x,y)$, $k=k(x,y)$, given by
\begin{equation}\label{ds sp}
    \begin{split}
    ds^2 &= -f\left( dt-\frac{\omega}{L} d [L\varphi] \right)^2 \\
    &\quad + L^2 f^{-1} \bigg( \frac{(x^2+1)(1-y^2)}{L^2} d[L\varphi]^2 \\ 
    &+(x^2+y^2) e^{2k} \left\{ \frac{dx^2}{x^2+1} +\frac{dy^2}{1-y^2} \right\} \bigg),
    \end{split}
\end{equation}
These coordinates are related to the Lewis-Papapetrou coordinates $(r, \theta)$ \cite{Cita:ExactSolutions}
by $y=\cos{\theta} \in [-1,1]$, $L x=r-l_1 \in \mathbb{R}$, being $l_1$ and $L$ are constants. 

Using , (\ref{4Potencial}), (\ref{ds sp}) and defining the potentials
\begin{subequations}\label{DefinicionPotenciales}
    \begin{equation}
        \Tilde{D}\chi=\frac{2f\kappa^2}{\rho} L(\frac{\omega}{L} DA_t +DA_\varphi),
    \end{equation}
    \begin{equation}
        \Tilde{D} \epsilon = \frac{f^2}{\rho} D\omega + \psi \Tilde{D}\chi,
    \end{equation}
    \begin{equation}
        \psi=2A_t,
    \end{equation}
\end{subequations} 
where the differential operators $D, \Tilde{D}$ are 
\begin{equation}
    D=
    \begin{bmatrix}
    \begin{array}{c}
    \partial_\rho \\
    \partial_z
    \end{array}
    \end{bmatrix},
    \qquad
    \Tilde{D}=
    \begin{bmatrix}
    \begin{array}{c}
    \partial_z \\
    -\partial_\rho
    \end{array}
    \end{bmatrix},
\end{equation}
the potential can be defined as $(Y^A)^T=\big[ f,\epsilon,\chi,\psi,\kappa \big]$ related to the gravitational, rotational, magnetic, electric and scalar potentials.

In terms of the $(Y^A)^T$ potentials, the fields equations (\ref{EcuacionesDeCampoOriginales}) take the form \footnote{For a comprehensive derivation, refer to \cite{Matos:2000ai}.}:

\begin{subequations}\label{EcuacionesDeCampoV2}
\begin{center}
    \text{\textbf{Klein-Gordon equation}}
    \begin{equation}
        D(\rho D\kappa ) - \frac{\rho }{\kappa} D\kappa^2+\frac{\rho \kappa^3 \alpha_0^2}{4 f \epsilon_0}  \left( D\psi^2-\frac{1}{\kappa^4} D\chi^2 \right) =0 \label{eq:KleinGordonV2} ,
    \end{equation}
    \text{\textbf{Maxwell equations}}
    \begin{multline}\label{Eq:Maxwell1V2}
        D (\rho D\psi)+\rho D\psi \left( \frac{2D\kappa}{\kappa} -\frac{Df}{f} \right) \\ -\frac{\rho}{f\kappa^2} (D\epsilon -\psi D\chi) D\chi =0 ,
    \end{multline}
    \begin{multline}\label{Eq:Maxwell2V2}
        D (\rho D\chi)-\rho D\chi \left( \frac{2D\kappa}{\kappa} +\frac{Df}{f} \right) \\ +\frac{\rho \kappa^2}{f} (D\epsilon -\psi D\chi) D\psi =0, 
    \end{multline}  
    \text{\textbf{Einstein equations}}
    \begin{multline}\label{Eq:Einstein1V2}
        D (\rho Df)+\frac{\rho}{f}\left( (D\epsilon-\psi D\chi )^2 -Df^2 \right) \\ -\frac{\rho \kappa^2}{2}\left( D\psi^2 +\frac{1}{\kappa^4} D\chi^2 \right) = 0 ,
    \end{multline}
    \begin{equation}\label{Eq:Einstein2V2}
        D (\rho (D\epsilon-\psi D\chi)) - \frac{2 \rho }{f} (D\epsilon-\psi D\chi) Df =0. 
    \end{equation}  
\end{center}
\end{subequations}

In \cite{Matos:2010pcd} a new method was proposed to solve the equations (\ref{EcuacionesDeCampoV2}), taking advantage of the fact that the potential space generated by $(Y^A)^T=\big[ f,\epsilon,\chi,\psi,\kappa \big]$ is conformally flat. This property allows the use of the ansatz $(Y^A)^T=(Y^A(\lambda))^T$, where the potential $\lambda=\lambda(x,y)$, satisfies Laplace's equation
\begin{equation}\label{Eq:LaplaceEnEsferoidales}
        \partial_x \{ (x^2+1)\partial_x \lambda \}+\partial_y \{(1-y^2) \partial_y \lambda \}=0,
\end{equation}
ensuring a space that simplifies the field equations. Using the fields $Y^A$ and the ansatz explained, in \cite{Matos:2010pcd} the following family of solutions is obtained
\begin{multline}\label{SegundaClaseSoluciones}
    f=f_0, \quad \kappa=\kappa_0 e^{\lambda}, \quad \psi= \frac{\sqrt{f_0}}{\kappa_0} e^{-\lambda} +\psi_0, \\ \quad \chi = \sqrt{f_0}\kappa_0 e^{\lambda} + \chi_0, \quad \epsilon= b_0,
\end{multline} 
where $\{f_0, \psi_0, \chi_0, \kappa_0, b_0 \}$ are integration constants.

Two solutions of equation (\ref{Eq:LaplaceEnEsferoidales}), is given by
\begin{subequations}\label{LambdaSoluciones}
   \begin{align}
       \lambda_5&=\lambda_0 \frac{x}{(x^2+y^2)} +m_0, \label{lambda5}\\
       \lambda_{N1}&=\lambda_0  xy+m_0, \label{lambda1N}
   \end{align} 
\end{subequations}
where $\lambda_0$ is an integration constant. The solution $\lambda_5$ has been taken from the appendix of the paper \cite{Matos:2010pcd}, while $\lambda_{N1}$ represents a new solution.

To get the metric functions $f,\omega,k$ and the electromagnetic potential $A_{\mu}$ back, we need to solve the following equations.
\begin{subequations}\label{FuncionesMetricas-xy}
\begin{center}
    \begin{equation}\label{EcuacionesDiferencialesOmega}
        \begin{bmatrix}
            \partial_x \\
            \partial_y \\
        \end{bmatrix} 
        \omega= \frac{L}{f^2}\Big( \epsilon_{,\lambda} -\psi \chi_{,\lambda} \Big)
        \begin{bmatrix}
            (1-y^2)\partial_y  \\
            -(x^2+1)\partial_x  
        \end{bmatrix}\lambda ,
    \end{equation}
    \begin{multline}\label{EcuacionesDiferencialesA3}
        \begin{bmatrix}
            \partial_x \\
            \partial_y \\
        \end{bmatrix} 
        A_\varphi = \frac{1}{2 f \kappa^2 }\Big( \chi_{,\lambda} \Big) 
        \begin{bmatrix}
            (1-y^2)\partial_y  \\
            -(x^2+1)\partial_x  
        \end{bmatrix} \lambda
        \\ -\frac{\omega}{2L} \Big( \psi_{,\lambda} \Big)
        \begin{bmatrix}
            \partial_x \\
            \partial_y \\
        \end{bmatrix} \lambda,
    \end{multline}
    \begin{multline}\label{EcuacionesDiferencialesk1}
        \partial_x k=k_{0}\frac{1-y^2}{x^2+y^2} \Big\{-2y (x^2+1)(\partial_x \lambda)(\partial_y \lambda) \\ +x\big[ (x^2+1)(\partial_x \lambda)^2-(1-y^2) (\partial_y \lambda)^2  \big]\Big\}, 
    \end{multline}
    \begin{multline}\label{EcuacionesDiferencialesk2}
        \partial_y k=k_{0}\frac{x^2+1}{x^2+y^2} \Big\{ 2x (1-y^2)(\partial_x \lambda)(\partial_y \lambda) \\ +y\big[ (x^2+1)(\partial_x \lambda)^2-(1-y^2) (\partial_y \lambda)^2  \big] \Big\}, 
    \end{multline}
\end{center}
\end{subequations}
where $k_0$ is an integration constant, and $F_{,\lambda}=\partial F / \partial \lambda$. These equations were obtained by inverting the definition of potentials (\ref{DefinicionPotenciales}). 
By substituting (\ref{SegundaClaseSoluciones}) and (\ref{LambdaSoluciones}) in (\ref{FuncionesMetricas-xy}) it is possible to obtain the corresponding solution.
The solution associated with $\lambda_5$ is the following
\begin{subequations}\label{SolucionLambda5}
\begin{align}
        f&=f_0=1,\\
        \omega &=-L\frac{ \lambda_0}{f_0 } \bigg( \frac{y(x^2+1)}{x^2+y^2} \bigg), \label{Omega Lambda5} \\
        A_\varphi &=-\frac{\sqrt{f_0}}{2 \kappa_0} \frac{\omega}{L} e^{-\lambda_5}, \label{A3 Lambda5} \\
        A_t&=\frac{\sqrt{f_0}}{2\kappa_0} e^{-\lambda_5},
\end{align}
\begin{multline}
    k = -k_{0} \lambda_0 ^2  \frac{(1-y^2)}{4(x^2+y^2)^4} \bigg(  -8x^2y^2(x^2+1)   \\  +[x^2+y^2]^2\big[(1-y^2) +2(x^2+y^2)\big] \bigg) . \label{k Lambda5} \\
\end{multline}
\end{subequations}
On the other hand, the one associated with $\lambda_{N1}$ is
\begin{subequations}\label{SolucionLambdaN1}
\begin{align}
        f&=f_0=1, \\
        \omega &=-\frac{ L \lambda_0 }{2f_0} \bigg(x^2(1-y^2)-y^2 \bigg), \label{Omega LambdaN1}\\
        k &= k_{0} \lambda_0 f_0 \omega/L, \label{k LambdaN1} \\
        A_\varphi &=-\frac{\sqrt{f_0}}{2 \kappa_0} \frac{\omega}{L} e^{-\lambda_{N1}}, \label{A3 LambdaN1} \\
        A_t&=\frac{\sqrt{f_0}}{2\kappa_0} e^{-\lambda_{N1}}.
\end{align}
\end{subequations}
Both solutions represent a rotating electromagnetic wormhole with no gravitational potential, $L$ is a constant related to the throat size of the wormhole.

When we select the scalar potential in this way
\begin{equation}\label{FormaDelPotencial}
    \phi=-\frac{\lambda}{\alpha_0},
\end{equation}
we substitute (\ref{FormaDelPotencial}), and (\ref{SolucionLambdaN1}) or (\ref{SolucionLambda5}) in (\ref{EcuacionesDeCampoOriginales}), which brings us to the constraint equation
\begin{equation}\label{EcuacionDeVerificacion}
    \alpha_0^2(4k_0+1)-4\epsilon_0=0 .
\end{equation}
This equation allows us to determine the values of $k_0$ in terms of the other constants.

 Using (\ref{Alpha0}) the possible values of $k_0$ are in Table [\ref{TablaValoresk}].

\begin{table}[b]
\caption{\label{TablaValoresk}%
Values of $k_0$}
\begin{ruledtabular}
\begin{tabular}{ccc}
\textbf{$\alpha_0^2$} &  
  \textbf{Dilatonic fiel ($\epsilon_0=1$)} & \textbf{Phantom fiel ($\epsilon_0=-1$)} \\
\colrule
1 & $3/4$ & $-5/4$ \\
3 & $1/12$ & $-7/12$ \\
4 & 0 & $-1/2$ \\
$n\geq 5$ & $(4-n)/4n$ & $-(4+n)/n$ 
\end{tabular}
\end{ruledtabular}
\end{table}

In what follows we will study the main feature of both solutions, showing their most important properties.

\section{Main Physical Properties of the Solutions}

To analyze the behavior of the spaces corresponding to solutions (\ref{SolucionLambda5}) and (\ref{SolucionLambdaN1}), the Ricci scalar $R=\tensor{g}{^{\alpha \beta}} \tensor{R}{_{\alpha \beta}}=\tensor{g}{^{\alpha \beta}} \tensor{R}{^\sigma}{_{\alpha \sigma \beta}}$ and the Kretschmann scalar $KN=\tensor{R}{^{ \mu \nu \alpha \beta}}\tensor{R}{}{_{ \mu \nu \alpha \beta}}$ will be examined. This will help to determine whether the space is asymptotically flat or contains singularities.

The invariants $R$ and $KN$ associated to the solution 1 ($\lambda_5$) are
\begin{subequations}
    \begin{multline}
        R=\frac{\lambda_0^2 (4k_0+1)}{2L^2 (x^2+y^2)^4}e^{-2k(x,y)} \big\{y^2 \\ +x^2(x^2+1-3y^2) \big\}, \label{Ricci para lambda5}
    \end{multline}
    \begin{align}
        KN=\frac{F_{5}(x,y)}{8L^{4}(x^2+y^2)^{12}} e^{-4k(x,y)}, \label{KN para lambda5}
    \end{align}
    
\end{subequations}

where, $F_{5}(x,y)$ is a polynomial of degree less than $(x^2+y^2)^{12}$.

The invariants $R$ and $KN$ associated to the solution 2 ($\lambda_{N1}$) are
\begin{subequations}
    \begin{equation}
        R= \lambda_0^2 \frac{(4k_0+1)}{2L^2} e^{-2k(x,y)},\label{Ricci para lambdaN1}
    \end{equation}
    \begin{align}
        KN=-\frac{F_{N1}(x,y)}{8L^4} e^{-4k(x,y)}, \label{KN para lambdaN1}
    \end{align}
\end{subequations}

where $F_{N1}(x,y)$ grows more slowly than the exponential.

By analyzing (\ref{Ricci para lambda5}) and (\ref{KN para lambda5}), we observe that a ring singularity exists in $x=y=0$ in the space associated with $\lambda_{5}$. The opposite occurs in the space associated with $\lambda_{N1}$, where the only singularity is governed by the exponential of (\ref{Ricci para lambdaN1}) and (\ref{KN para lambdaN1}). That is, for a dilatonic-type scalar field, the singularity is located at $x=\pm \infty$, but if a phantom scalar field is chosen, the space is regular at every point.

By closely examining (\ref{Ricci para lambda5}) and (\ref{KN para lambda5}), we observe that the space corresponding to $\lambda_{5}$ is asymptotically flat regardless of the choice of scalar field, phantom or dilatonic. This is because, 

\begin{widetext}
\begin{equation*}
    \lim\limits_{r\rightarrow \pm \infty } k(x,y)=\lim\limits_{x \rightarrow \pm \infty } k(x,y)=0\qquad \Rightarrow \qquad \lim\limits_{x \rightarrow \pm \infty } R=0 \qquad \& \lim\limits_{x \rightarrow \pm \infty } KN=0,
\end{equation*}
\end{widetext}

since the degree of the numerator of $R$ and $KN$ is lower than that of the denominator, and the exponential does not play a role since the denominator of the metric function $k(x,y)$ is greater than the numerator, and the exponential tends to 1. 

In contrast, when analyzing (\ref{Ricci para lambdaN1}) and (\ref{KN para lambdaN1}), the space is asymptotically flat only when choosing a phantom scalar field, because the exponential $e^{-2k(x,y)}$, dictates the behavior of the limit of $R$ and $KN$, i.e., the function metric $k(x,y)$ is important because

\begin{widetext}
\begin{equation*}
    \lim\limits_{x \rightarrow \pm \infty } k(x,y)= 
            \begin{cases} 
                +\infty &  \text{ \textbf{if}} \quad k_0 < 0 \\
                -\infty &  \text{ \textbf{if}} \quad k_0 > 0
            \end{cases} \qquad \Rightarrow \qquad
    \lim\limits_{x \rightarrow \pm \infty } \{R, KN\}= 
            \begin{cases} 
                0 &  \text{ \textbf{if}} \quad k_0 < 0 \\
                 \infty &  \text{ \textbf{if}} \quad k_0 > 0
            \end{cases}.      
\end{equation*}
\end{widetext}

\section{Null Energy Condition}

The condition to determine whether the scalar field is associated with normal or exotic matter will involve the null energy condition 
\begin{equation*}
    T_{\mu \nu} l^{\mu} l^{\nu} \geq 0, 
\end{equation*}
for all $l^{\mu}$, where $\tensor{T}{_{\mu \nu}}$ is the energy-momentum tensor and $l^{\mu}$ is the null vector that satisfies $l^{\mu}l_{\mu}=0$.

We can obtain a more tractable representation in the comoving reference frame using the procedure proposed by Morris and Thorne in their paper \cite{Morris:1988cz}.
To derive the representation of the tensors in the comoving reference frame ($\hat{\mathbb{O}}$), the following transformation matrix is necessary:
\begin{widetext}
\begin{equation}\label{MatrizDeTransformacionComovil}
    \mathbb{M}_{comoving}^{T}=\begin{bmatrix}
        \begin{array}{cccc}
             1/\sqrt{-\tensor{g}{_{t t } }} & 0 & 0 & 0 \\
             0 & 1/\sqrt{\tensor{g}{_{x x } }} & 0 & 0 \\
             0 & 0 & 1/\sqrt{\tensor{g}{_{y y } }} & 0 \\
             \frac{-\tensor{g}{_{\varphi t}}/\tensor{g}{_{t t}} }{\sqrt{\tensor{g}{_{\varphi \varphi}}-(\tensor{g}{_{\varphi t}})^2/\tensor{g}{_{t t}}}} & 0 & 0 & \frac{1}{\sqrt{\tensor{g}{_{\varphi \varphi}}-(\tensor{g}{_{\varphi t}})^2/\tensor{g}{_{t t}}}}
        \end{array}
    \end{bmatrix}.
\end{equation}
\end{widetext}

So, the basis vector in $\hat{\mathbb{O}}$ takes the form
\begin{equation}\label{RelacionDeBasesComovilMetricaTesis}
    \begin{bmatrix}
    \begin{array}{c}
     \tensor{e}{_{\hat{t}}} \\
     \tensor{e}{_{\hat{x}}} \\
     \tensor{e}{_{\hat{y}}} \\
     \tensor{e}{_{\hat{\varphi}}}
    \end{array}
    \end{bmatrix}^{T}
    =\mathbb{M}_{comoving}^{T}
    \begin{bmatrix}
    \begin{array}{c}
     \tensor{e}{_{t}} \\
     \tensor{e}{_{x}} \\
     \tensor{e}{_{y}} \\
     \tensor{e}{_{\varphi}}
    \end{array}
    \end{bmatrix}^{T},
\end{equation}
and the Ricci tensor
\begin{equation*}
    \tensor{R}{_{\hat{\mu} \hat{\nu}}}=\mathbb{M}_{comoving} \cdot \{R_{\mu \nu} \}\cdot (\mathbb{M}_{comoving})^{T}.
\end{equation*}

In this frame, a significant relationship is derived, considering $l_{\mu}=e_{\hat{t}}+e_{\hat{x}}$

\begin{multline}\label{Eq: De la condicion de energía}
    T_{\hat{\mu} \hat{\nu}} l^{\hat{\mu}} l^{\hat{\nu}}= T_{\hat{t} \hat{t}}+ T_{\hat{x} \hat{x}}=\rho-\varrho \\ =\frac{1}{8 \pi}\bigg( R_{\hat{t} \hat{t}}+ R_{\hat{x} \hat{x}} \bigg) \geq 0,
\end{multline}
where is used natural unities ($c=1=G$), $R_{\hat{\mu}\hat{\mu}}$ is the $\mu$-th Ricci element in the comovil reference frame, $\rho$ is the density and $-\varrho$ the pressure.

For the second class of solutions (\ref{SegundaClaseSoluciones}), we find that the density satisfies the following condition:

\begin{multline}\label{Densidad 2da familia de soluciones}
    8\pi \rho =R_{\hat{t}\hat{t}}= \frac{e^{-2k}}{2L^4 (x^2+1) (1-y^2) (x^2 + y^2)} \\ \bigg( (1-y^2) \omega_{,y}^2 + (x^2+1) \omega_{,x}^2 \bigg) \geq 0,
\end{multline}

due to the fact that there are only quadratic terms in (\ref{Densidad 2da familia de soluciones}), and $A_{,z}=\partial A/ \partial z$.

On the other hand, the expression for the pressure is

\begin{multline}\label{Presion 2da familia de soluciones}
        -8\pi \varrho= \tensor{R}{_{\hat{x} \hat{x}}}= -\frac{e^{-2k}}{L^2 (x^2 + y^2)} \bigg( (1-y^2) k_{,yy} \\+ (x^2+1) k_{,xx} -2y k_{,y} - \frac{\omega_{,x}^2}{2L^2 (1-y^2)} \bigg) ,
\end{multline}

we can observe that if there is no scalar field and no rotation, pressure will not exist.

By substituting the corresponding metric functions for $\lambda_{5}$ (\ref{SolucionLambda5}) and $\lambda_{N1}$ (\ref{SolucionLambdaN1}) in (\ref{Presion 2da familia de soluciones}), the following equalities are obtained:

NEC associated to the solution 1 ($\lambda_{5}$)
\begin{multline}\label{rho - varrho Lambda5}
    \rho-\varrho=\frac{\lambda_0^2 (4 k_0+1) e^{-2k}}{2L^2 (x^2+y^2)^5}   \bigg( x^6+x^4 \left(1-2 y^2\right) \\+y^4 +\frac{x^2 y^2 \left( 4 k_0 \left(y^2-2\right)-7 y^2+6\right)}{4 k_0+1}\bigg),
\end{multline}

NEC associated to the solution 2 ($\lambda_{N1}$)
\begin{multline}\label{rho - varrho LambdaN1}
    \rho-\varrho=\frac{\lambda_0^2 e^{-2k}}{2L^2 (x^2+y^2)}  \Bigg(2x^2 \\+y^2(4k_0+1+x^2(4k_0-1)) \Bigg),
\end{multline}

To determine the satisfaction of the null energy condition, it is necessary to carefully analyze (\ref{rho - varrho Lambda5}) and (\ref{rho - varrho LambdaN1}).

We begin by noting that both equations contain terms with even powers. Specifically, $y\in [-1,1] \quad \Rightarrow \quad y^{2n} \in [0,1]$, appear in one case, $x \in (-\infty, \infty) \quad \Rightarrow \quad x^{2n} \in [0,\infty)$.  In the case of the exponential function, we observe that for (\ref{k Lambda5}) and (\ref{k LambdaN1}), again there are terms raised to an even power, then $e^{-2k} \in [0,\infty )$.

Thus, the sign will be determined by the terms enclosed in parentheses and in brackets in (\ref{rho - varrho Lambda5}) and (\ref{rho - varrho LambdaN1}) respectively.

For better understanding, we will evaluate both expressions at $y=0$ and $y=1$, obtaining:

\begin{subequations}
\begin{itemize}
    \item $(\lambda_{5}) \qquad y=0$:
    \begin{equation}\label{rho - varrho _Lambda5-y=0}
        \rho-\varrho = \frac{\lambda_0^{2} e^{-2k(x,0)}}{2L^2 x^6} (x^2+1) \bigg(  4k_0+1 \bigg),
    \end{equation}

    \item $(\lambda_{5}) \qquad y=1$:
    \begin{equation}\label{rho - varrho _Lambda5-y=1}
        \rho-\varrho =  \frac{(\lambda_0 [x^2-1])^2}{2L^2 (x^2+1)^4} \bigg( 4k_0+1 \bigg),
    \end{equation}
\end{itemize}
\end{subequations}

\begin{subequations}
\begin{itemize}
    \item $(\lambda_{N1}) \qquad y=0$:
    \begin{equation}\label{rho - varrho _LambdaN1-y=0}
        \rho-\varrho = \frac{e^{-2k(x,0)}}{2L^2} \lambda_0^2.
    \end{equation}

    \item $(\lambda_{N1}) \qquad y=1$:
    \begin{equation}\label{rho - varrho _LambdaN1-y=1}
        \rho-\varrho = \frac{\lambda_0^2 e^{-2k(x,1)}}{2L^2}  \bigg( 4k_0+1 \bigg) .
    \end{equation}
\end{itemize}
\end{subequations}

For all cases except (\ref{rho - varrho _LambdaN1-y=0}), the expression within the brackets will determine the satisfaction of the null energy condition (NEC). Upon examination of Table [\ref{TablaValoresk}], we can conclude that, as expected, the NEC is satisfied only when a dilatonic scalar field is chosen.
\section{Wormhole Geometry}

To thoroughly investigate the geometry of the wormhole, we will consider the hypersurface specified by $t=t_0$ and $y=y_0$ within the solution obtained, i.e. $f=1$.

\begin{multline}\label{ds Hipersuperficie}
    ds^2 =  L^2\frac{(x^2+y_0^2) e^{2k(x,y_0)}}{x^2+1}  dx^2 \\+\left[ L^2 (x^2+1)(1-y_0^2) -\omega(x,y_0)^2 \right] d\varphi^2.
\end{multline}

This hypersurface is embedded in a cylindrical space that can be parameterized using the given coordinates $(x,y)$.

\begin{align}
    ds^2&=d\rho^2+dz^2+\rho^2d\varphi^2 \nonumber \\
    &=\left\{ \left( \frac{d\rho}{dx} \right)^2 +\left( \frac{dz}{dx} \right)^2 \right\}dx^2+\rho(x,y_0)^2 d\varphi^2. \label{CilindricasParametrizadasHipersup}
\end{align}

Subsequent to the embedding process, we derived the following equations, which can only be solved using numerical methods, given the initial condition $z(0)=0$ at the point corresponding to the throat of the wormhole.

\begin{subequations}\label{Ec de la GoemtriaHipersup}
    \begin{align}
        &\rho(x,y_0)^2= L^2 (x^2+1)(1-y_0^2) -\omega(x,y_0)^2  , \label{rho(x} \\
        &\left( \frac{d\rho}{dx} \right)^2 +\left( \frac{dz}{dx} \right)^2= L^2\frac{(x^2+y_0^2)}{x^2+1} e^{2k(x,y_0)}. \label{EcDif rhoZ(x)}
    \end{align}
\end{subequations}

By numerically solving and obtaining $\rho (x)$ and $z(x)$, we can plot $z(\rho)$ against a given value of $y=\cos{\theta}=y_0$.

\subsection{Solution \texorpdfstring{$\lambda_{5}$}{lambda-5}}

\begin{figure}[b]
    \centering
    \begin{minipage}{0.45\textwidth}
    \centering
        \includegraphics[width=\textwidth]{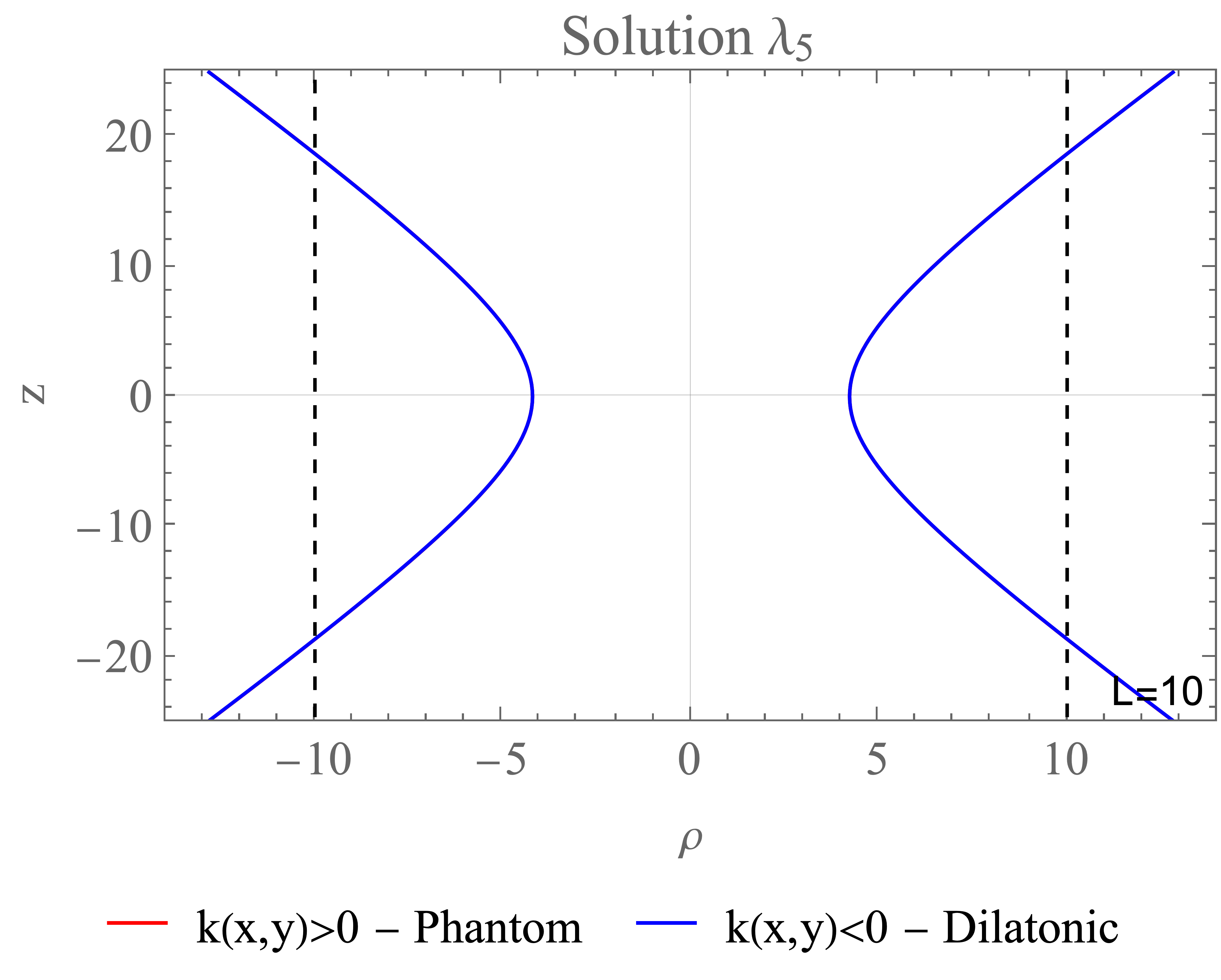}
        \subcaption{A comparative analysis of the curves $z (\rho)$ for \textit{phantom scalar field} and \textit{dilatonic scalar field} using the solution associated with $\lambda_5$. The respective values used are $\lambda_0=1/10$, $k_0=\{-5/4,3/4 \}$, $y_0=0.9$, $f=1$, and $L=10$.}
        \label{fig:SolucionLambda5-ComparacionPFyDF}
    \end{minipage}
    \hfill
    \begin{minipage}{0.45\textwidth}
    \centering
        \includegraphics[width=\textwidth]{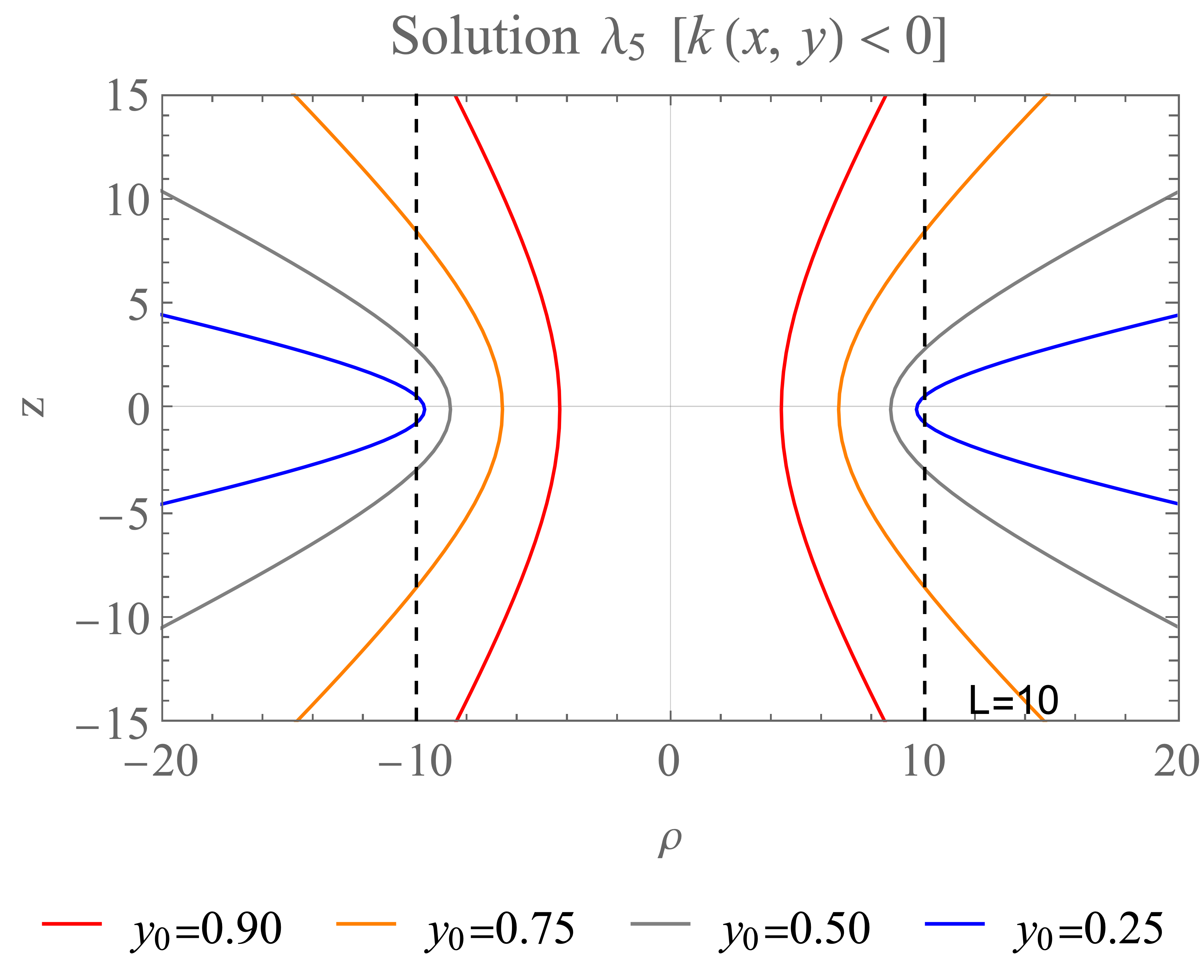}
        \subcaption{The geometry of the wormhole pertaining to the $\lambda_5$ solution is examined. The values used include $\lambda_0=1/10^2$, $f=1$, $k_0=1/12$, and $L=10$, along with various $y_0$ to demonstrate the behavior of the curves $z (\rho)$. In addition, two black dotted lines were incorporated at $z(0)=\pm L$. \textit{Note: $k(x,y)<0$ was used to focus on the study of phenomena that satisfy NEC.}}
        \label{fig:Lambda5-VariasYs}
    \end{minipage}
    \caption{\label{fig:Lambda5Ambas}}
\end{figure}

The geometry of the wormhole corresponding to $\lambda_{5}$ for a specific $y_0$ is illustrated in Figure [\ref{fig:SolucionLambda5-ComparacionPFyDF}]. As shown, the choice of scalar field, whether phantom or dilatonic, is irrelevant to the formation and form of the wormhole, i.e. its geometry. The geometry with a dilatonic scalar field is plotted in blue, and a phantom-type field in red. In addition, the value $L=10$ is graphical, related to the throat size of the wormhole, and is represented by a black dashed line. This is determined by the sign of the integration constant $k_0$ (See Table [\ref{TablaValoresk}]), which dictates the sign of the metric function $k(x,y)$.

Figure [\ref{fig:Lambda5-VariasYs}] presents the geometry of a dilatonic wormhole for different values of $y_0$, illustrating the behavior of the wormhole as a function of the polar entry angle. As $y_0$ approaches zero (that is, $\theta \rightarrow \pi /2$), the throat approaches to $L$, with its smallest size near the poles. This behavior is consistent with that of a phantom-type wormhole.

\subsection{Solution \texorpdfstring{$\lambda_{N1}$}{lambda-N1}}

\begin{figure}[b]
    \centering
    \begin{minipage}{0.45\textwidth}
    \centering
        \includegraphics[width=\textwidth]{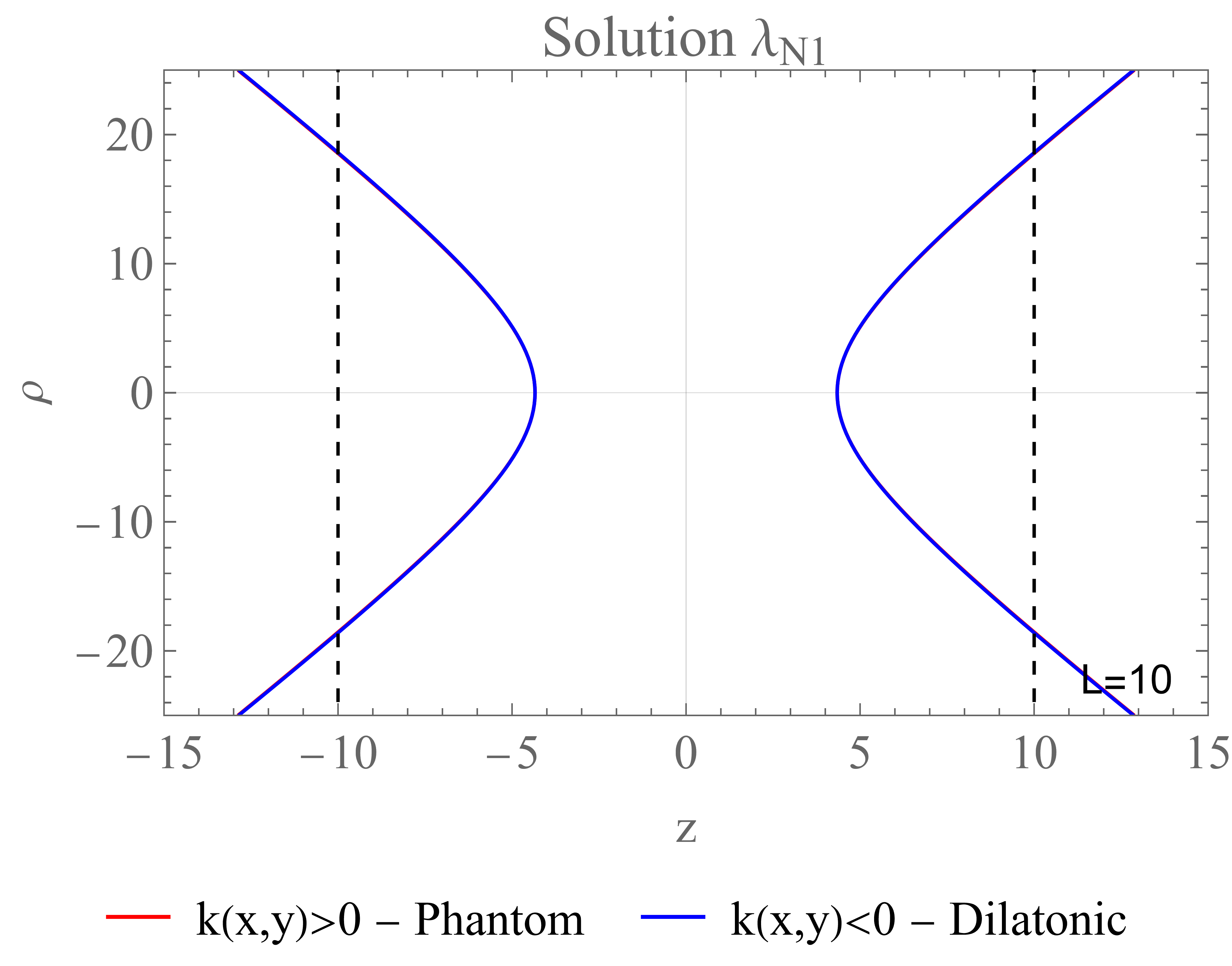}
        \subcaption{A comparative analysis of the curves $z (\rho)$ for a \textit{phantom scalar field} and a \textit{dilatonic scalar field} employing the solution associated with $\lambda_{N1}$. The parameters used are $\lambda_0=1/10$, $y_0=0.9$, $f=1$, and $k_0=\{-5/4,3/4 \}$ respectively, $L=10$.}
        \label{fig:SolucioN1-ComparacionPFyDF}
    \end{minipage}
    \hfill
    \begin{minipage}{0.45\textwidth}
    \centering
        \includegraphics[width=\textwidth]{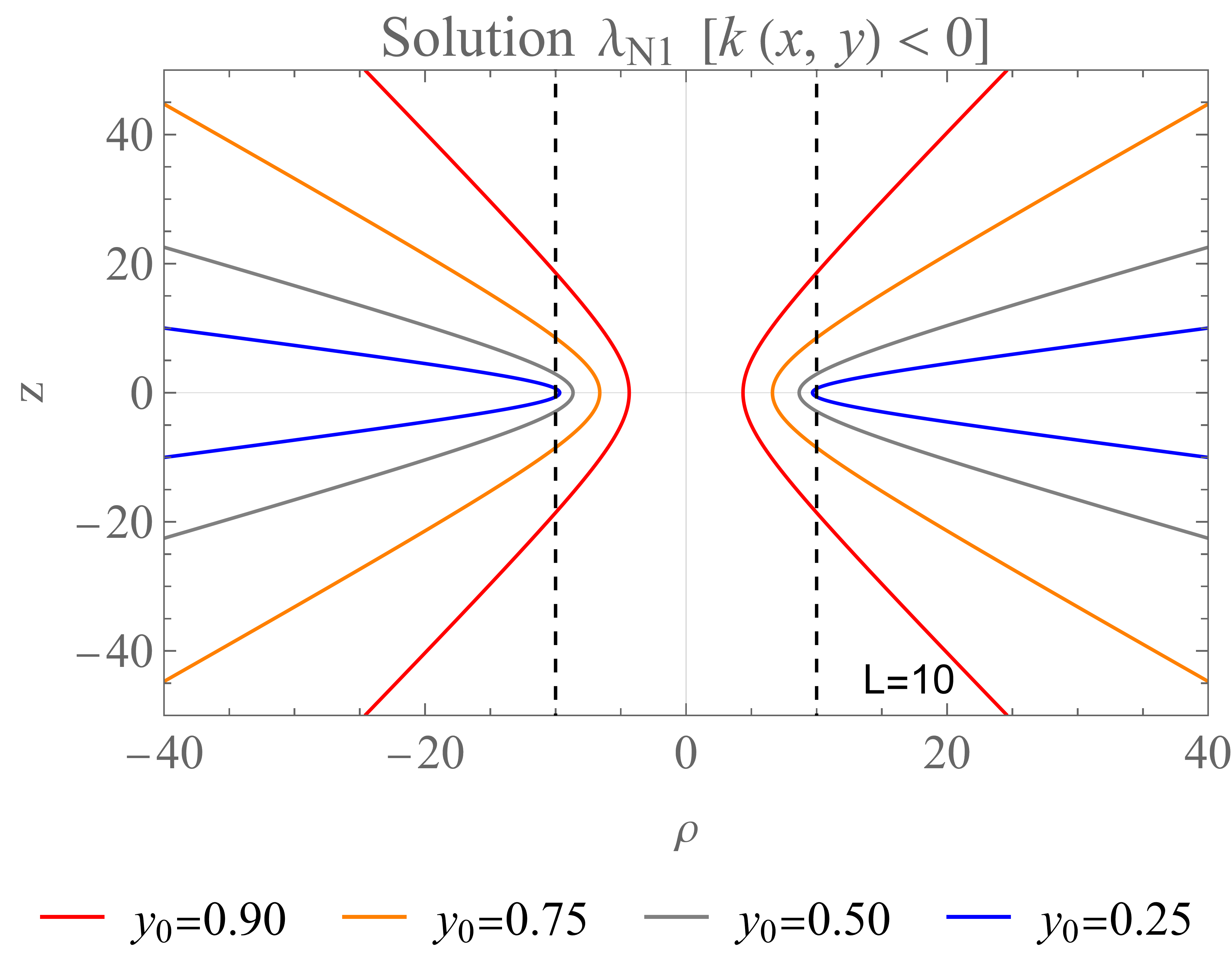}
        \subcaption{The geometry of the wormhole corresponding to the $\lambda_{N1}$ solution. The parameters used include $\lambda_0=1/10^3$, $f=1$, $k_0=1/12$, $L=10$, and the variation of $y_0$ to demonstrate the behavior of the curves $z (\rho)$. Additionally, two black dotted lines were included in $z(0)=\pm L$, the size of the wormhole throt.}
        \label{fig:SolucionN1-VariasYs}
    \end{minipage}
    \caption{\label{fig:LambdaN1Ambas}}
\end{figure}

The same analysis was applied to the wormhole corresponding to $\lambda_{N1}$. Again, the formation of the wormhole is independent of the scalar field type. In Figure [\ref{fig:SolucioN1-ComparacionPFyDF}] the dilatonic wormhole is shown in blue, the phantom-type is shown in red, and the value of $L$ is plotted with a black dashed line.

In Figure [\ref{fig:SolucionN1-VariasYs}], the geometry of the dilatonic wormhole is shown for various values of $y_0$. Once again, near the equatorial plane, the throat approaches $L$, while near the poles, it reaches its minimum value.

\section{Tidal Forces}

To analyze the tidal forces, the calculations will be performed in the astronaut's reference frame for simplicity. The spheroidal coordinates $(x,y)$ will continue to be used for the derivations, and the results will later be interpreted using the Boyer-Lindquist coordinates $(r,\theta)$. 

It will be assumed that in the reference frame of the astronaut $\hat{\overline{\mathbb{O}}}$, its 4-velocity is purely temporal $\vartheta^{\hat{\overline{\mu}}}=(1,0,0,0)$, ensuring the orthogonality of the 4-acceleration and 4-velocity $W^{\hat{\overline{\mu}}} \vartheta_{\hat{\overline{\mu}}}=0 \quad \Rightarrow \quad W^{\hat{\overline{\mu}}}=(0,\overline{a},0,0)$. 
Furthermore, the astronaut is considered to traverse the compact object radially, without any rotation, to simplify the scenario. Thus, using the geodesic deviation equation, which relates the geometry of the wormhole to the tidal forces experienced by the traveler
\begin{equation}\label{Tidal Forces}
    \Delta a^{\hat{\overline{\mu}}}=-c^2 \tensor{R}{^{\hat{\overline{\mu}}}}{_{\hat{\overline{\alpha}} \hspace{0.2em} \hat{\overline{\beta}} \hspace{0.2em} \hat{\overline{\sigma}} }} \tensor{\vartheta}{^{\hat{\overline{\alpha}}}} \tensor{\xi}{^{\hat{\overline{\beta}}}} \tensor{\vartheta}{^{\hat{\overline{\sigma}}}},
\end{equation}
we can determine the danger zones, where $\xi^{\hat{\overline{\mu}}}$ is the 4-distance from the astronaut's head to his feet, and $\tensor{R}{^{\hat{\overline{\mu}}}}{_{\hat{\overline{\alpha}} \hspace{0.2em} \hat{\overline{\beta}} \hspace{0.2em} \hat{\overline{\sigma}} }}$ are the components of the Riemann Tensor in the reference frame of the astronaut.

To obtain the components of the Riemann tensor in the reference frame of the astronaut $\hat{\overline{\mathbb{O}}}$, we use special relativity tools. The process is as follows: starting from an arbitrary reference frame $\mathbb{O}$, the Riemann tensor is transformed into the comoving reference frame $\hat{\mathbb{O}}$ with (\ref{MatrizDeTransformacionComovil}). It is then transformed from the comoving reference frame to the astronaut's reference frame using the matrix
\begin{equation}\label{MatrizTransformaciónRelEsp}
    \mathbb{M}_{RE}=
    \begin{bmatrix}
        \begin{array}{cccc}
            \beta & -\gamma \beta & 0 & 0 \\
            -\gamma \beta & \beta & 0 & 0 \\
            0 & 0 & 1 & 0 \\
            0 & 0 & 0 & 1 
        \end{array}
    \end{bmatrix},
\end{equation}
where $\beta= \vartheta /c $, $\vartheta$ is the speed of $\hat{\overline{\mathbb{O}}}$ with respect of $\hat{\mathbb{O}}$, and $\gamma= 1/\sqrt{1-\beta^2}$.

By doing this and considering the relevant assumptions previously mentioned, we arrive at the following

\begin{subequations}\label{Riemman en SR del astronauta}
\begin{equation}\label{Rrtrt en SR del astronauta}
    \tensor{R}{_{ \hat{\overline{r}} \hspace{0.2em} \hat{\overline{t}} \hspace{0.2em} \hat{\overline{r}} \hspace{0.2em} \hat{\overline{t}}}}  =\tensor{R}{_{ \hat{r} \hat{t} \hat{r} \hat{t} }},
\end{equation}
\begin{equation}\label{R2t2t en SR del astronauta}
    \tensor{R}{_{ \hat{\overline{\theta}} \hspace{0.2em} \hat{\overline{t}} \hspace{0.2em} \hat{\overline{\theta}} \hspace{0.2em} \hat{\overline{t}}}} = \gamma^2 \tensor{R}{_{ \hat{\theta} \hat{t} \hat{\theta} \hat{t} }} + \gamma^2 \beta^2 \tensor{R}{_{ \hat{\theta} \hat{r} \hat{\theta} \hat{r} }},
\end{equation}
\begin{equation}\label{R3t3t en SR del astronauta}
    \tensor{R}{_{ \hat{\overline{\varphi}} \hspace{0.2em} \hat{\overline{t}} \hspace{0.2em} \hat{\overline{\varphi}} \hspace{0.2em} \hat{\overline{t}}}}=\gamma^2 \tensor{R}{_{ \hat{\varphi} \hat{t} \hat{\varphi} \hat{t} }} + \gamma^2 \beta^2 \tensor{R}{_{ \hat{\varphi} \hat{r} \hat{\varphi} \hat{r} }}.
\end{equation}
\end{subequations}

If we consider the height of the astronaut $|\xi| \approx 2 m$ and the maximum gravity the human body can withstand as that of Earth $g_{Earth}$, the following inequality must hold

\begin{equation}\label{CondicionFuerzasDeMarea}
    |\tensor{R}{_{\hat{\overline{\mu}} \hspace{0.2em} \hat{\overline{t}} \hspace{0.2em} \hat{\overline{\mu}} \hspace{0.2em} \hat{\overline{t}} }}| \leq \frac{g_{Earth}}{2m * c^2}\approx (10^5 Km)^{-2}.
\end{equation}

When considering that the astronaut travels at relativistic speeds ($\beta \approx 0 \quad \Rightarrow \quad \gamma \approx 1$), then $\tensor{R}{_{\hat{\overline{\mu}} \hspace{0.2em} \hat{\overline{t}} \hspace{0.2em} \hat{\overline{\mu}} \hspace{0.2em} \hat{\overline{t}} }} \approx \tensor{R}{_{ \hat{\mu} \hat{t} \hat{\mu} \hat{t} }}$. If relativistic speeds are taken into account, it is necessary to consider the terms $\tensor{R}{_{ \hat{\mu} \hat{r} \hat{\mu} \hat{r} }} $. The absolute value of the components of the Riemann tensor will be plotted at $\hat{\mathbb{O}}$, and each will be analyzed to determine the optimal entry angles to the WH.

\subsection{Riemann components in \texorpdfstring{$\hat{\mathbb{O}}$}{O-hat} corresponding to \texorpdfstring{$\lambda_{5}$}{lambda-5}}

To gain a clearer understanding of the tidal forces within the wormhole, it is essential to examine the components of the Riemann tensor. Upon calculating these components, the following expressions are obtained

\begin{subequations}\label{Riemman en SR del astronauta - Lambda5}

\begin{equation}\label{R2121-Lambda5}
    \tensor{R}{_{ \hat{x} \hat{t} \hat{x} \hat{t} }}=\frac{e^{-2k(x,y)}}{L^2 (x^2+y^2)^5} \big( \lambda_0 x y \big)^2 (1-y^2),
\end{equation}
\begin{equation}\label{R3131-Lambda5}
    \tensor{R}{_{ \hat{y} \hat{t} \hat{y} \hat{t} }} =\frac{e^{-2k(x,y)}}{4 L^2 (x^2+y^2)^5 } \lambda_0^2 (x^2+1)(x^2-y^2)^2 ,
\end{equation}
\begin{multline}\label{R4141-Lambda5}
    \tensor{R}{_{ \hat{\varphi} \hat{t} \hat{\varphi} \hat{t} }} =\frac{e^{-2k(x,y)}}{4 L^2 (x^2+y^2)^4} \lambda_0^2 \Bigg( x^4+y^2\\+x^2(1-3y^2) \Bigg),
\end{multline}
\begin{equation}\label{R3232-Lambda5}
    \tensor{R}{_{ \hat{y} \hat{x} \hat{y} \hat{x} }} = 4 k_0 \tensor{R}{_{ \hat{\varphi} \hat{t} \hat{\varphi} \hat{t} }},
\end{equation}
\begin{multline}\label{R4242-Lambda5}
    \tensor{R}{_{ \hat{\varphi} \hat{x} \hat{\varphi} \hat{x} }} =  \frac{e^{-2k(x,y)}}{L^2 (x^2+y^2)^5} \lambda_0^2 \Bigg\{ 3x^2 y^2 (1-y^2) \\ +k_0 \bigg( x^6+y^4+x^4(1-2y^2)+x^2 y^2 (5y^2-6)\bigg) \Bigg\}.
\end{multline}
\end{subequations}

\begin{figure*}
    \centering
    \begin{minipage}{0.27\textheight}
        \centering
        \begin{subfigure}{\textwidth}
            \includegraphics[width=\textwidth]{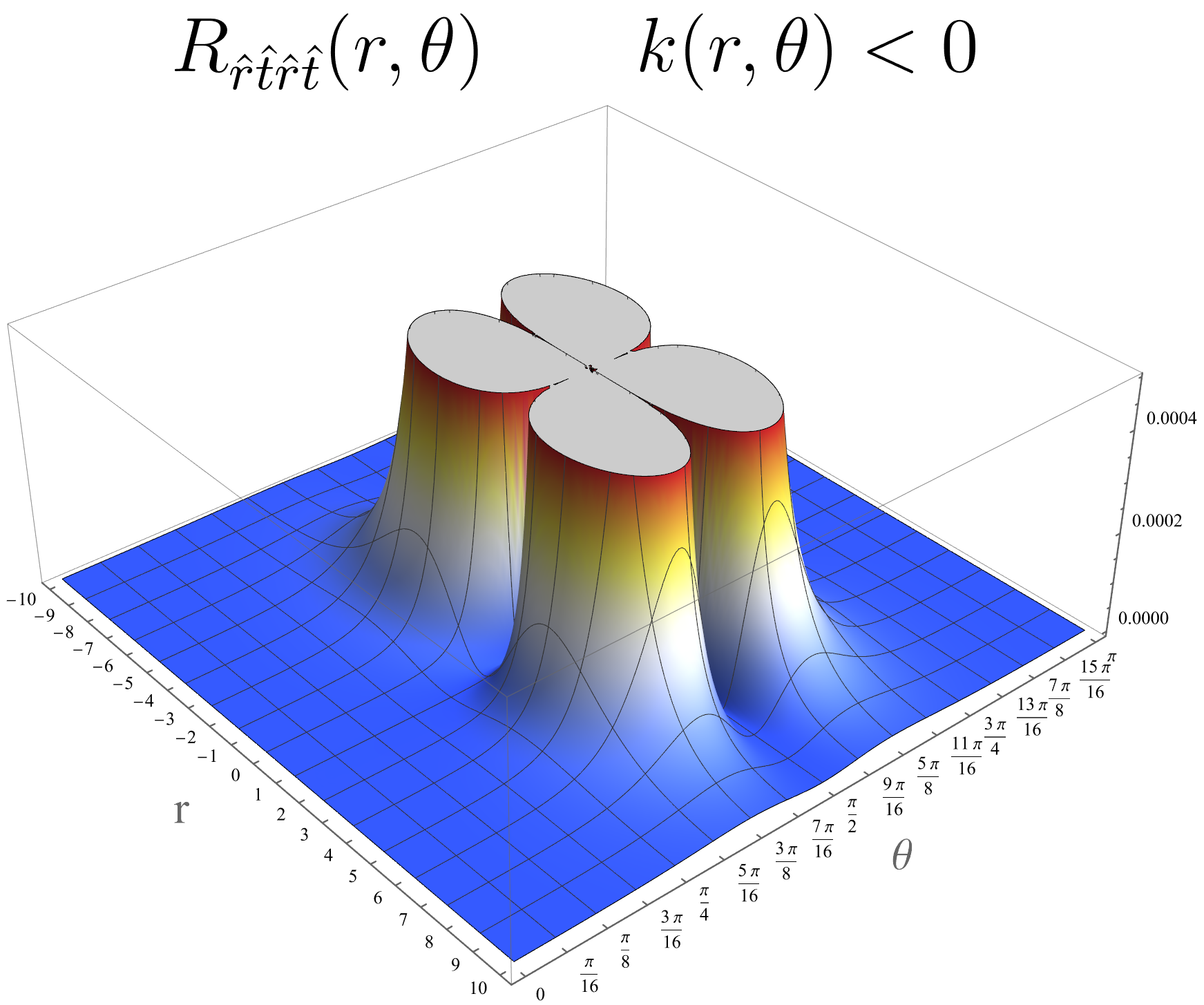}
            \caption{}
            \label{fig:L5_Dilaton-3D-RDDDD2121}
        \end{subfigure}
    \end{minipage}%
    \hfill
    \begin{minipage}{0.27\textheight}
        \centering
        \begin{subfigure}{\textwidth}
            \includegraphics[width=\textwidth]{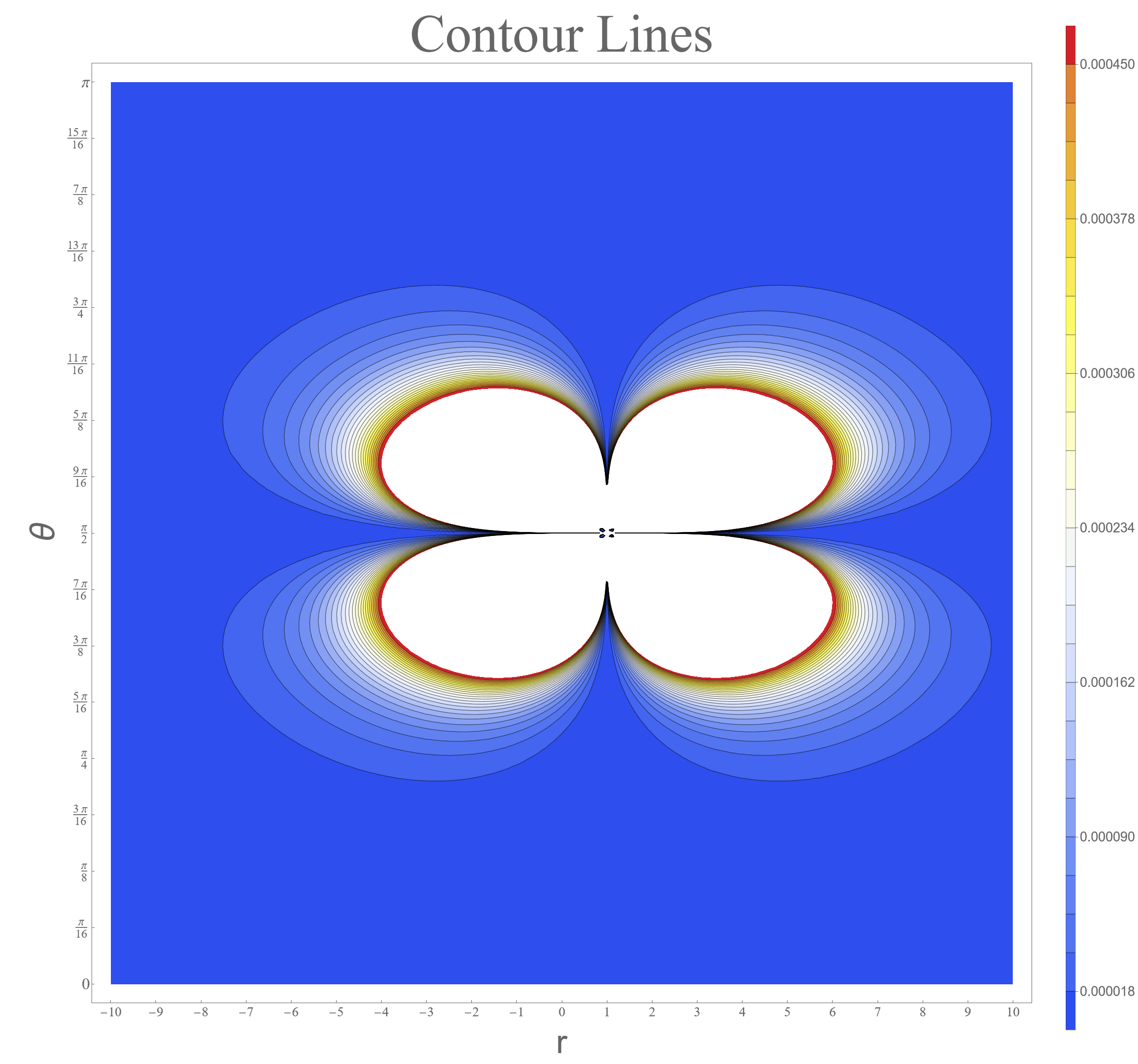}
            \caption{}
            \label{fig:L5_Dilaton-CurvNivel-RDDDD2121}
        \end{subfigure}
    \end{minipage}
    \centering
    \begin{minipage}{0.27\textheight}
        \centering
        \begin{subfigure}{\textwidth}
            \includegraphics[width=\textwidth]{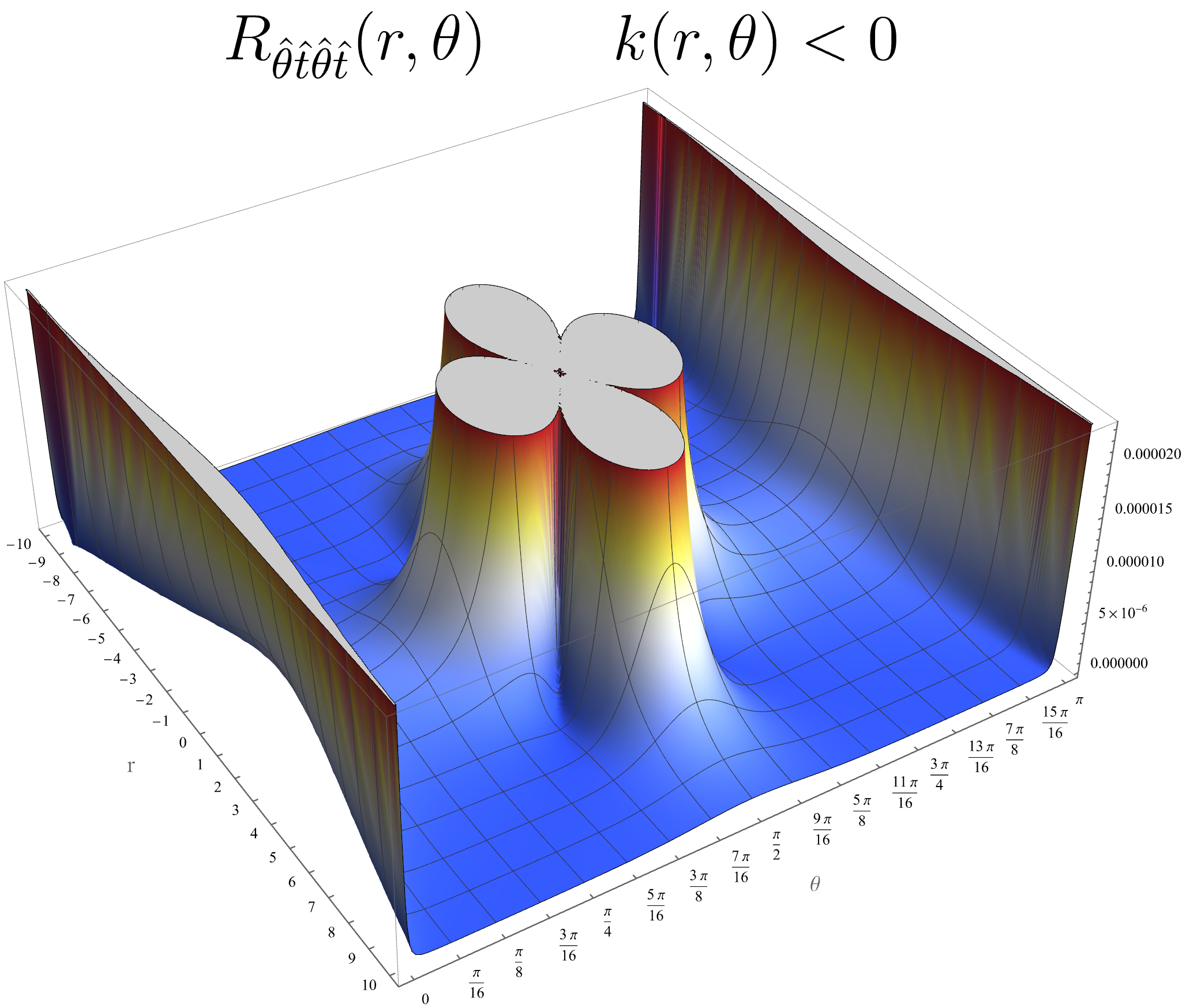}
            \caption{}
            \label{fig:L5_Dilaton-3D-RDDDD3131}
        \end{subfigure}
    \end{minipage}%
    \hfill
    \begin{minipage}{0.27\textheight}
        \centering
        \begin{subfigure}{\textwidth}
            \includegraphics[width=\textwidth]{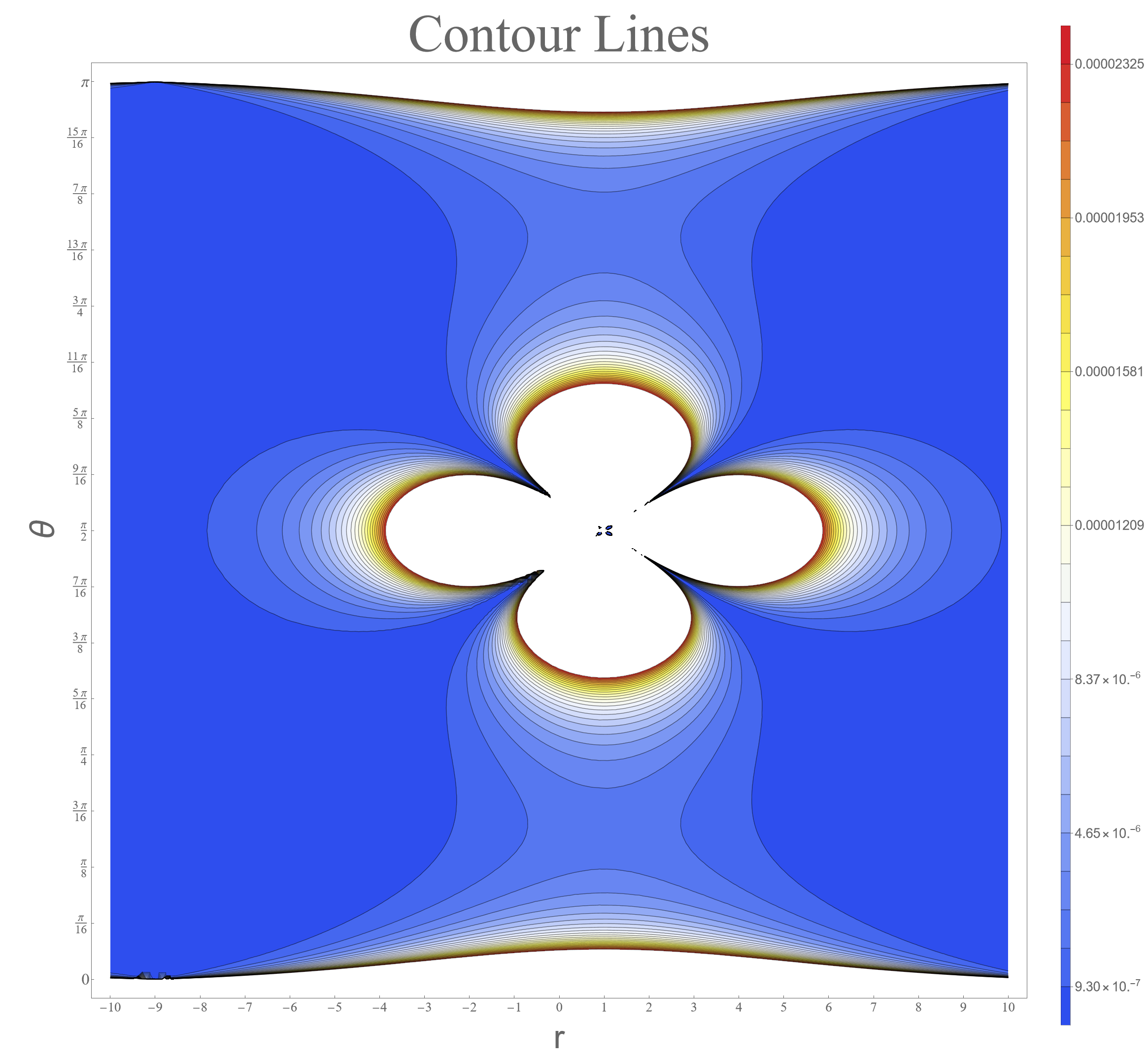}
            \caption{}
            \label{fig:L5_Dilaton-CurvNivel-RDDDD3131}
        \end{subfigure}
    \end{minipage}
    \centering
    \begin{minipage}{0.27\textheight}
        \centering
        \begin{subfigure}{\textwidth}
            \includegraphics[width=\textwidth]{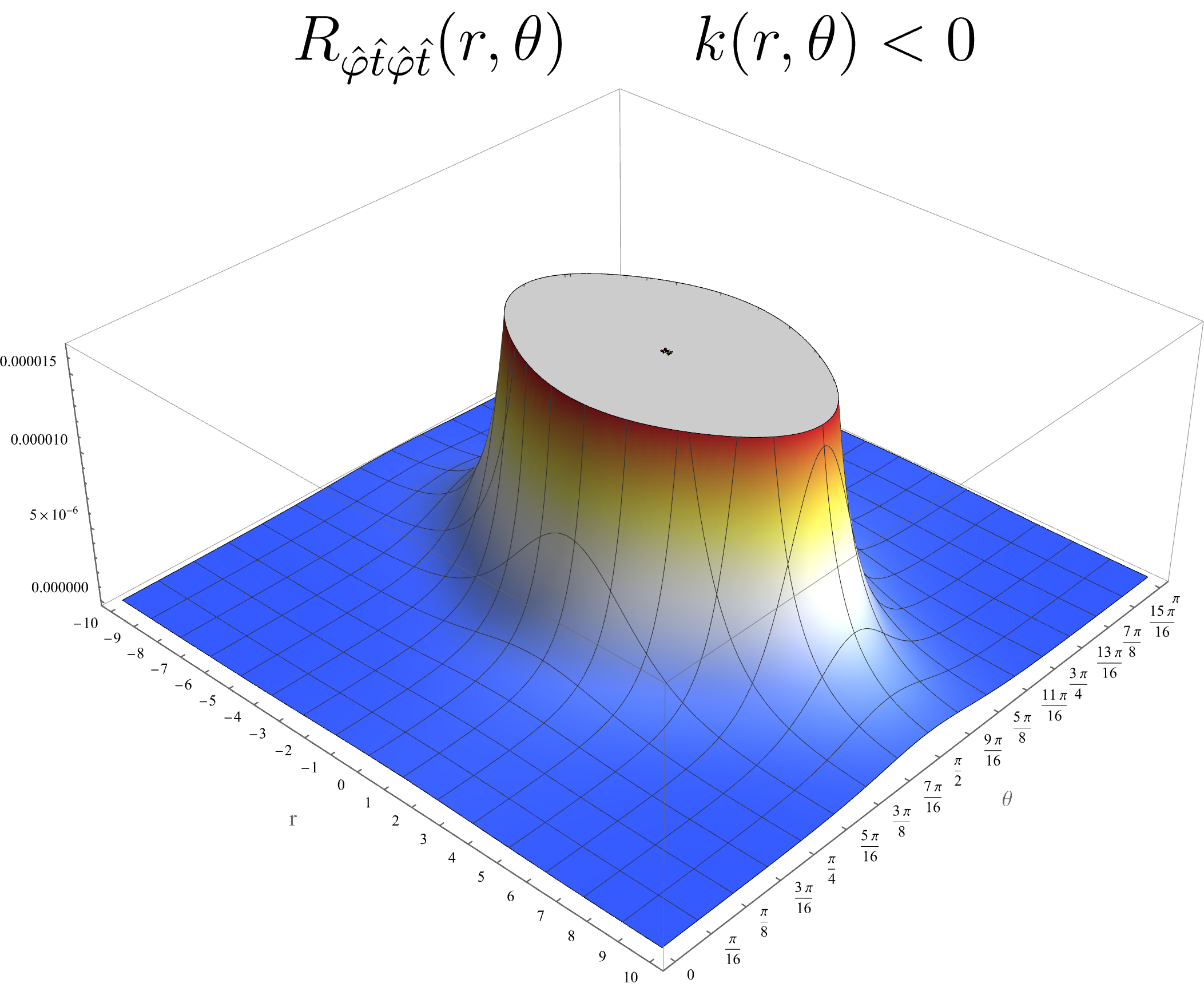}
            \caption{}
            \label{fig:L5_Dilaton-3D-RDDDD4141}
        \end{subfigure}
    \end{minipage}%
    \hfill
    \begin{minipage}{0.27\textheight}
        \centering
        \begin{subfigure}{\textwidth}
            \includegraphics[width=\textwidth]{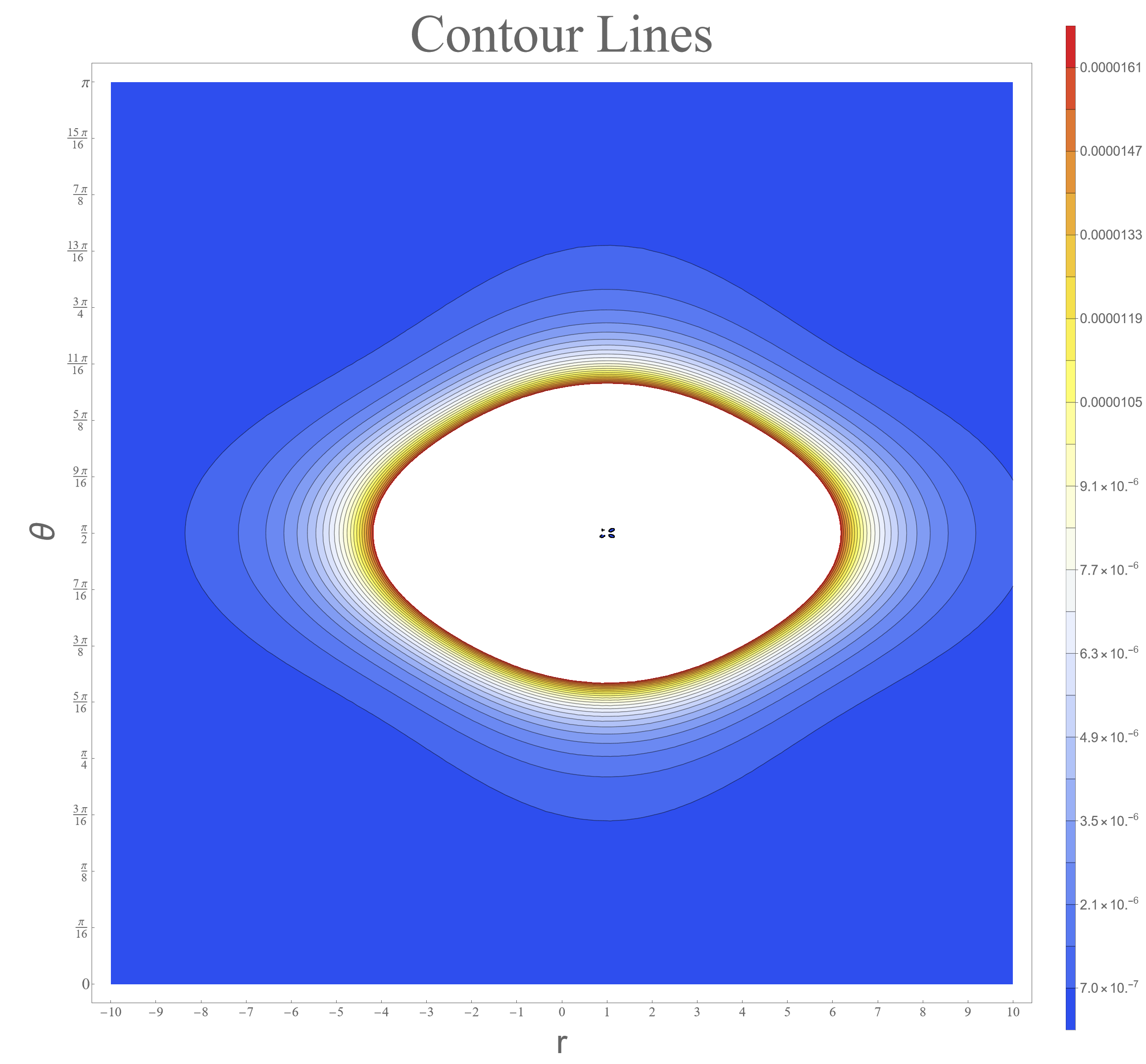}
            \caption{}
            \label{fig:L5_Dilaton-CurvNivel-RDDDD4141}
        \end{subfigure}
    \end{minipage}
    \caption{ 
    Graphical representations of the elements $\tensor{R}{_{ \hat{\mu} \hat{t} \hat{\mu} \hat{t} }} , \quad \hat{\mu}=\hat{r},\hat{\theta},\hat{\varphi}$ associated with $\lambda_5$. The parameters $\lambda_0=1/10^2$, $k_0=1/12$, $l_1=1$ and $L=10$ were used, and the coordinates ($r,\theta$) were selected to improve the physical visualization. The plots on the left depict the 3-D surface generated by $\tensor{R}{_{ \hat{\mu} \hat{t} \hat{\mu} \hat{t} }} , \quad \hat{\mu}=\hat{r},\hat{\theta},\hat{\varphi}$, while the plots on the right illustrate the contour lines. Both sides employ color schemes relevant to \textbf{high values (red color)} and \textbf{low values (blue color)}.}.
    \label{fig:RsNoCruzadasD_5}
\end{figure*}

Through the graphical representation of each component, the behavior of tidal forces is elucidated more effectively in Boyer-Linquist coordinates ($r,\theta$). Figure [\ref{fig:RsNoCruzadasD_5}] illustrates the primary components under the assumption of non-relativistic velocities and dilatonic scalar field ($k_0=1/12$). The graphs presented on the left depict three-dimensional surfaces, whereas those on the right-hand side represent contour lines, providing a more explicit interpretation of the magnitude of each graph. These visualizations employ a color gradient, where blue signifies the lowest magnitude values and red denotes the higher magnitude values of $\tensor{R}{_{ \hat{\mu} \hat{t} \hat{\mu} \hat{t} }}$ in the axis $z$. Furthermore, regions rendered in white indicate areas where the magnitude increases drastically, signifying an exponential growth in the associated Riemann component.

The $x,y$ axis corresponds to the radius $r$ and the polar angle $\theta$. Analysis of graphs [\ref{fig:L5_Dilaton-3D-RDDDD2121}] and [\ref{fig:L5_Dilaton-3D-RDDDD3131}] demonstrates that traversing the wormhole's center is achievable by varying the parameter $\theta$, thereby allowing navigation through it. In contrast, graph [\ref{fig:L5_Dilaton-3D-RDDDD4141}] reveals that access to the wormhole center is prohibited. However, all graphs consistently show the absence of tidal forces at the center of the wormhole $r=l_1=1, \theta\approx \pi /2$ that could potentially cause its destruction. It is also crucial to consider the magnitude: the components $\tensor{R}{_{ \hat{\theta} \hat{t} \hat{\theta} \hat{t} }}$ and $\tensor{R}{_{ \hat{\varphi} \hat{t} \hat{\varphi} \hat{t} }}$ are marginally smaller in magnitude relative to $\tensor{R}{_{ \hat{r} \hat{t} \hat{r} \hat{t} }}$, which is the predominant component for non-relativistic velocities, notwithstanding a lack of rigorous precision.

\begin{figure*}
    \begin{minipage}{0.27\textheight}
        \centering
        \begin{subfigure}{\textwidth}
            \includegraphics[width=\textwidth]{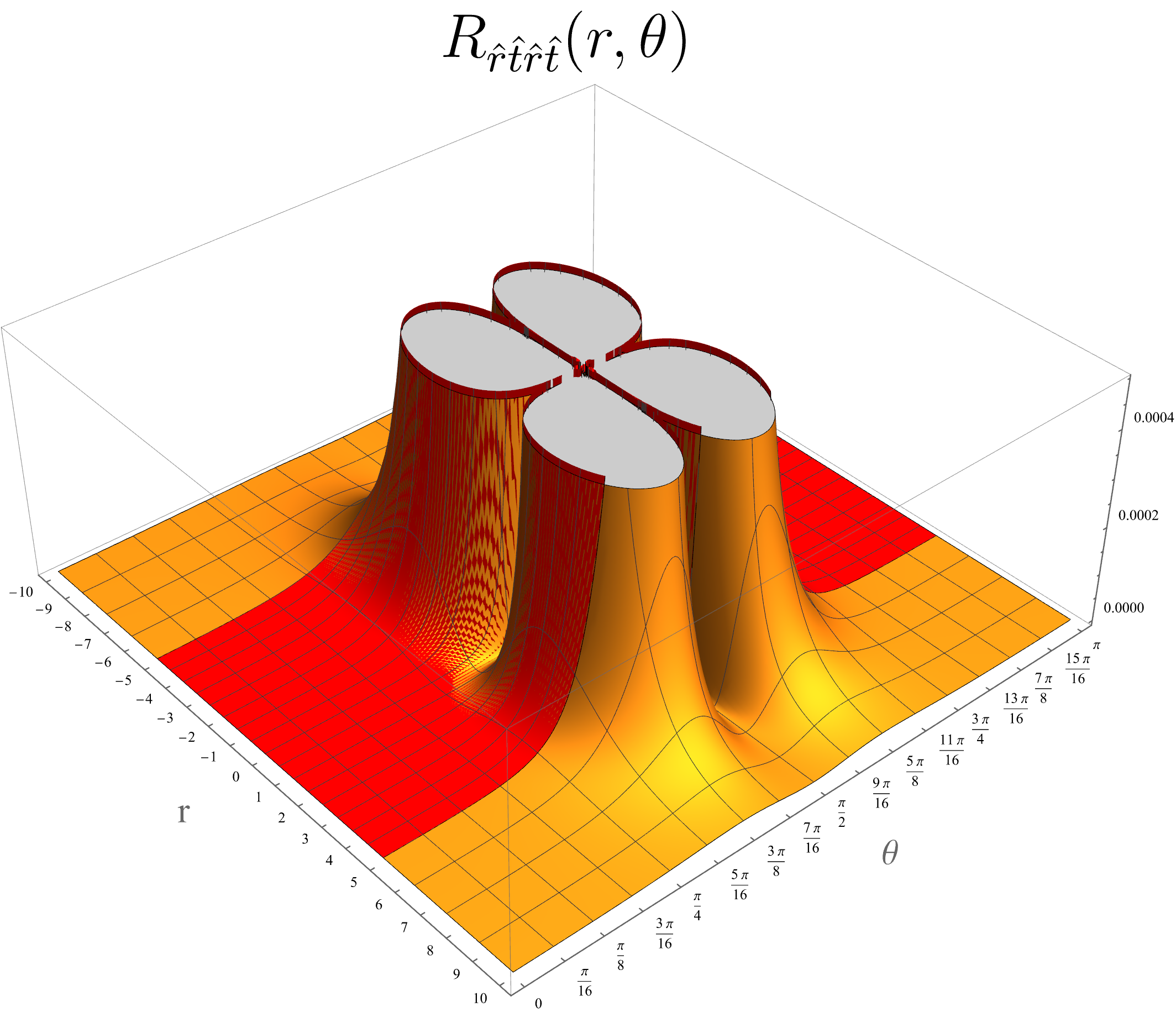}
            \caption{}
            \label{fig:L5_DilatonFantasma-3D Comparacion-RDDDD2121}
        \end{subfigure}
    \end{minipage}%
    \hfill
    \begin{minipage}{0.27\textheight}
        \centering
        \begin{subfigure}{\textwidth}
            \includegraphics[width=\textwidth]{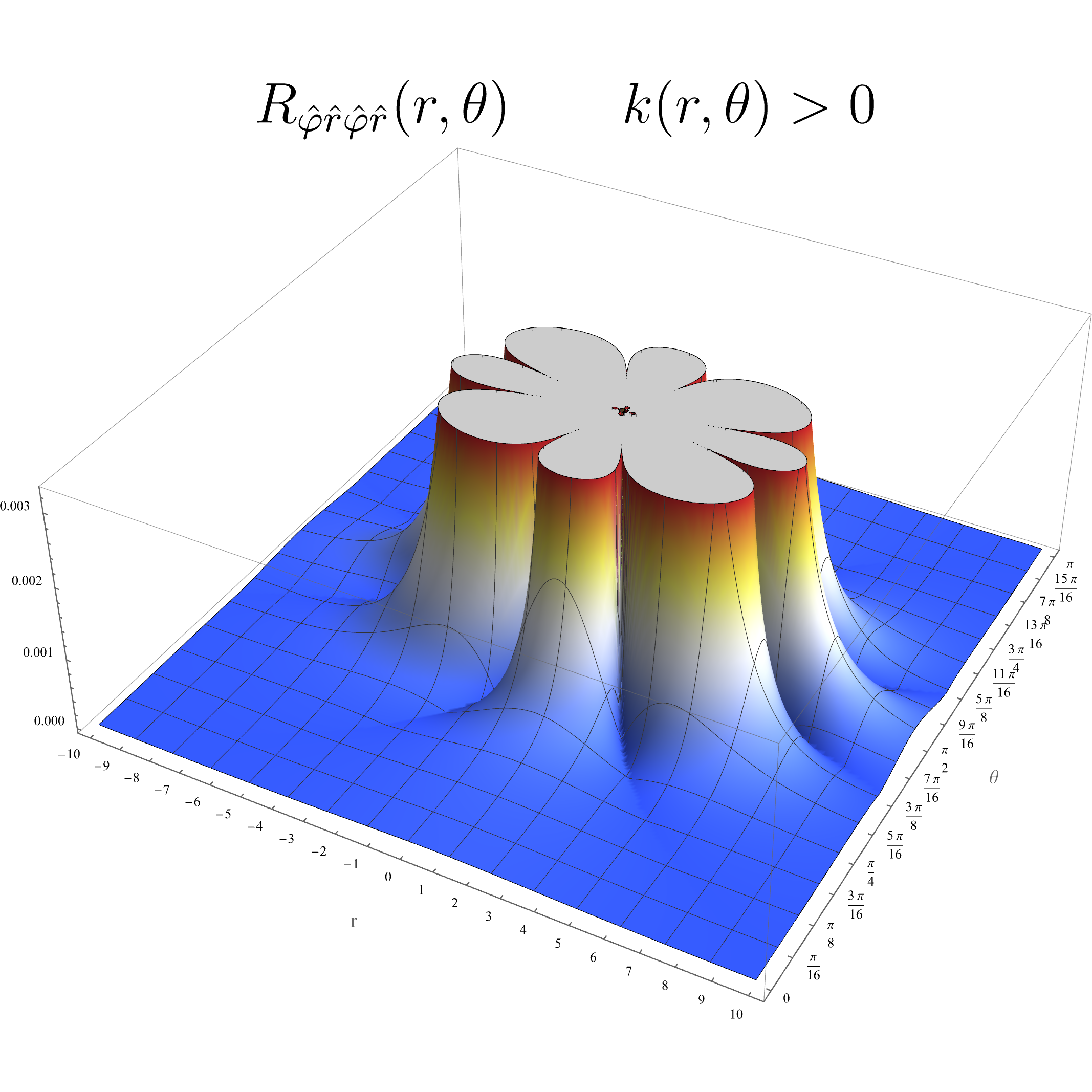}
            \caption{}
            \label{fig:L5_Fantasma-3D-RDDDD4242}
        \end{subfigure}
    \end{minipage}
    \centering
    \begin{minipage}{0.27\textheight}
        \centering
        \begin{subfigure}{\textwidth}
            \includegraphics[width=\textwidth]{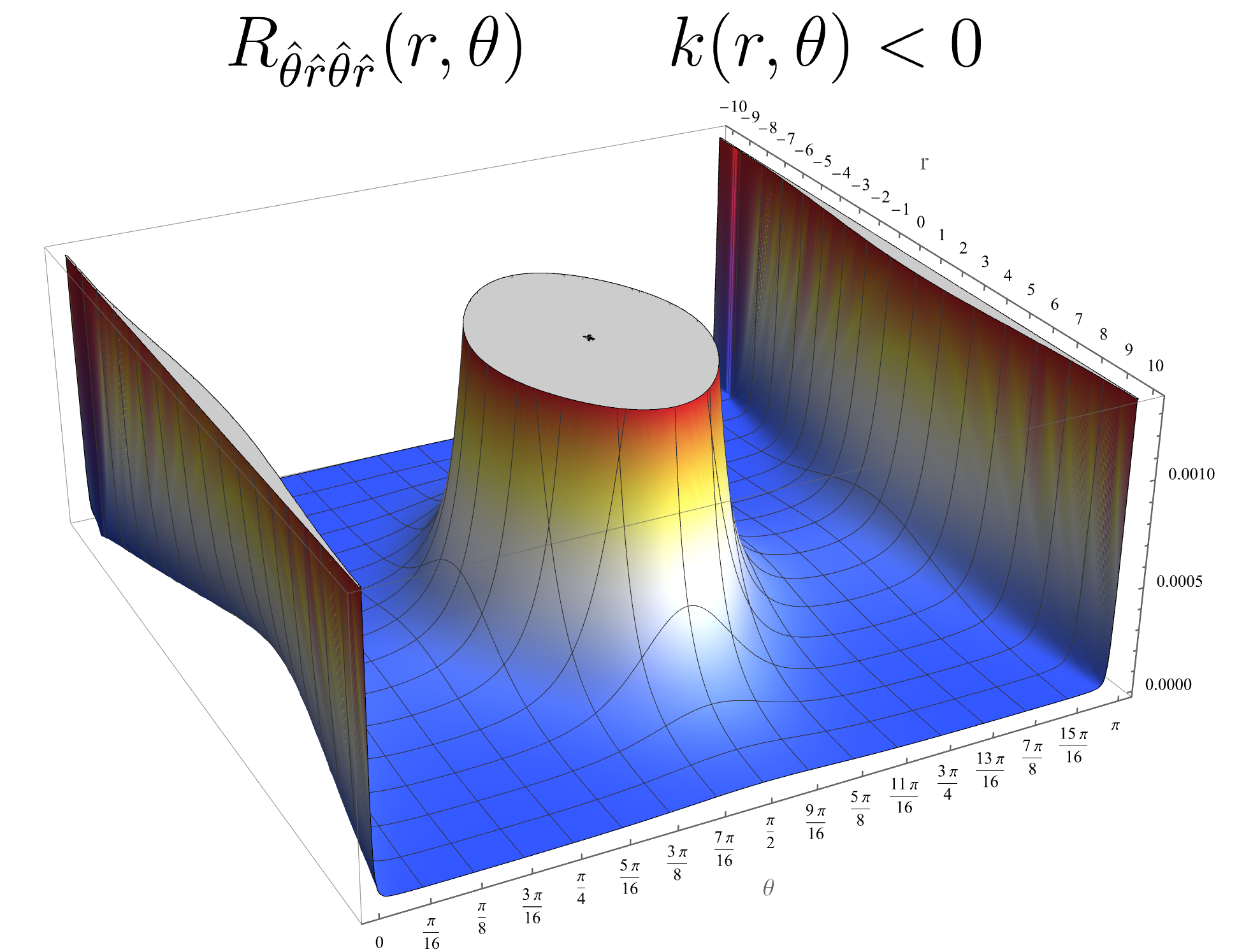}
            \caption{}
            \label{fig:L5_Dilaton-3D-RDDDD3232}
        \end{subfigure}
    \end{minipage}%
    \hfill
    \begin{minipage}{0.27\textheight}
        \centering
        \begin{subfigure}{\textwidth}
            \includegraphics[width=\textwidth]{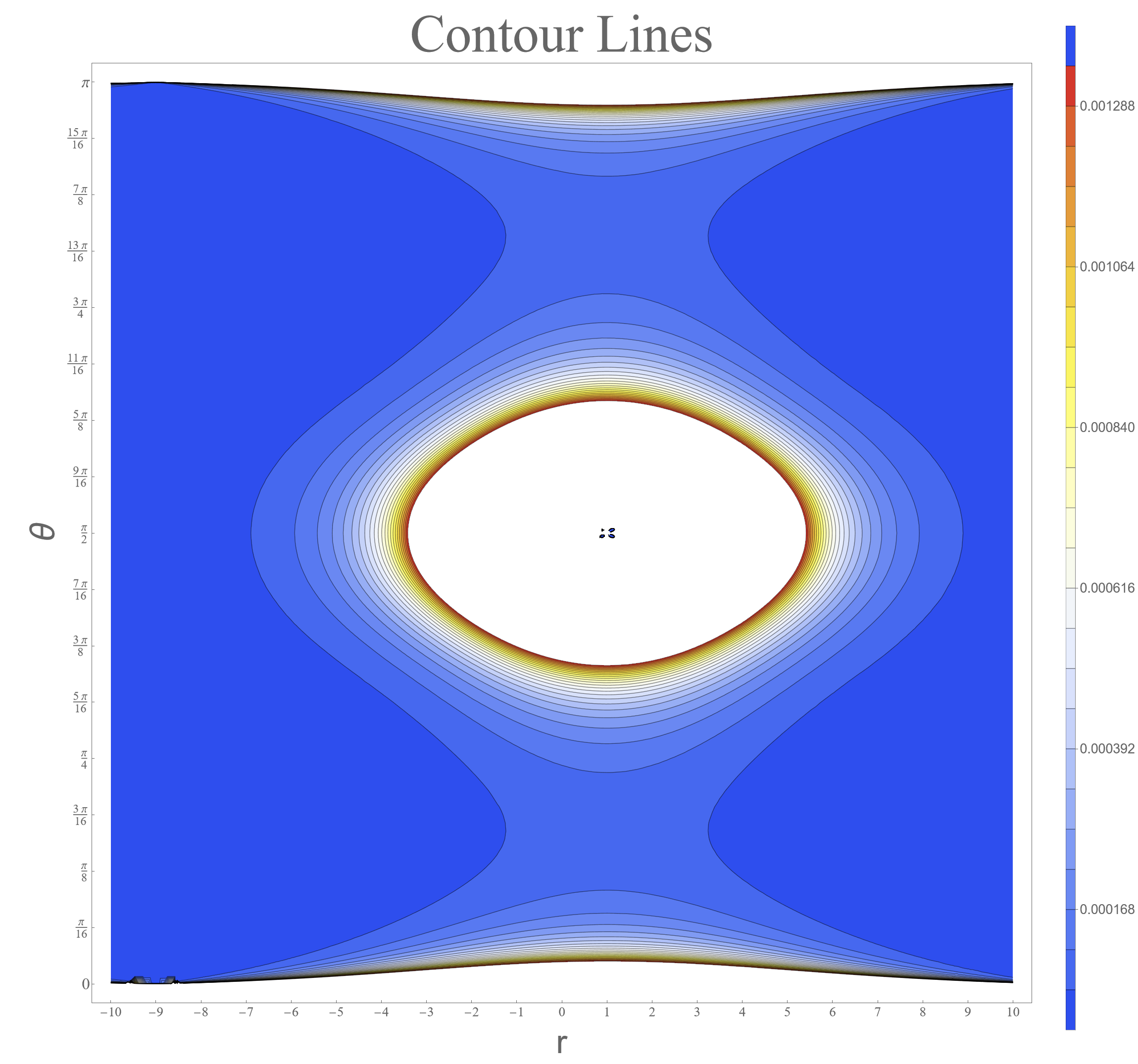}
            \caption{}
            \label{fig:L5_Dilaton-CurvNivel-RDDDD3232}
        \end{subfigure}
    \end{minipage}
    \centering
    \begin{minipage}{0.27\textheight}
        \centering
        \begin{subfigure}{\textwidth}
            \includegraphics[width=\textwidth]{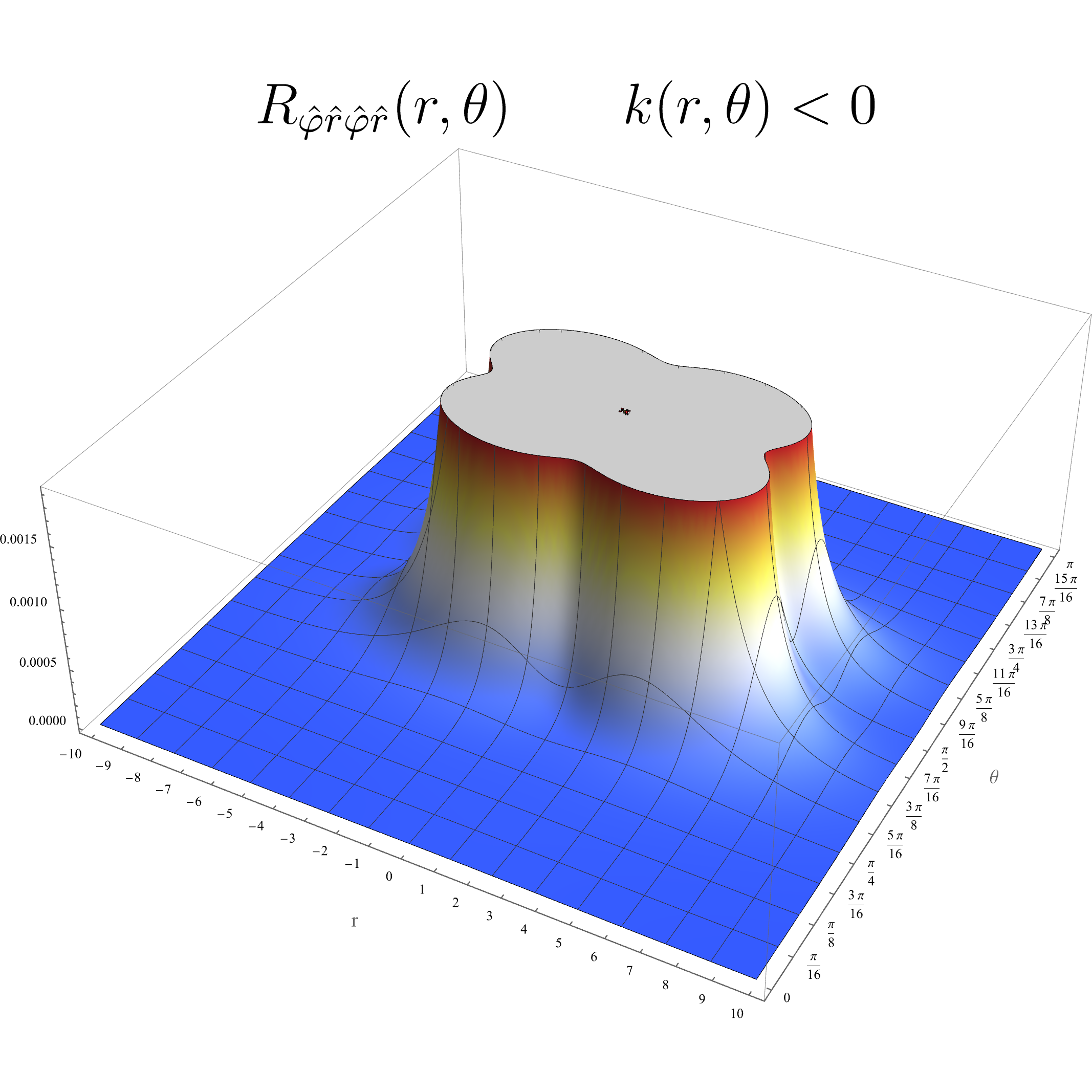}
            \caption{}
            \label{fig:L5_Dilaton-3D-RDDDD4242}
        \end{subfigure}
    \end{minipage}%
    \hfill
    \begin{minipage}{0.27\textheight}
        \centering
        \begin{subfigure}{\textwidth}
            \includegraphics[width=\textwidth]{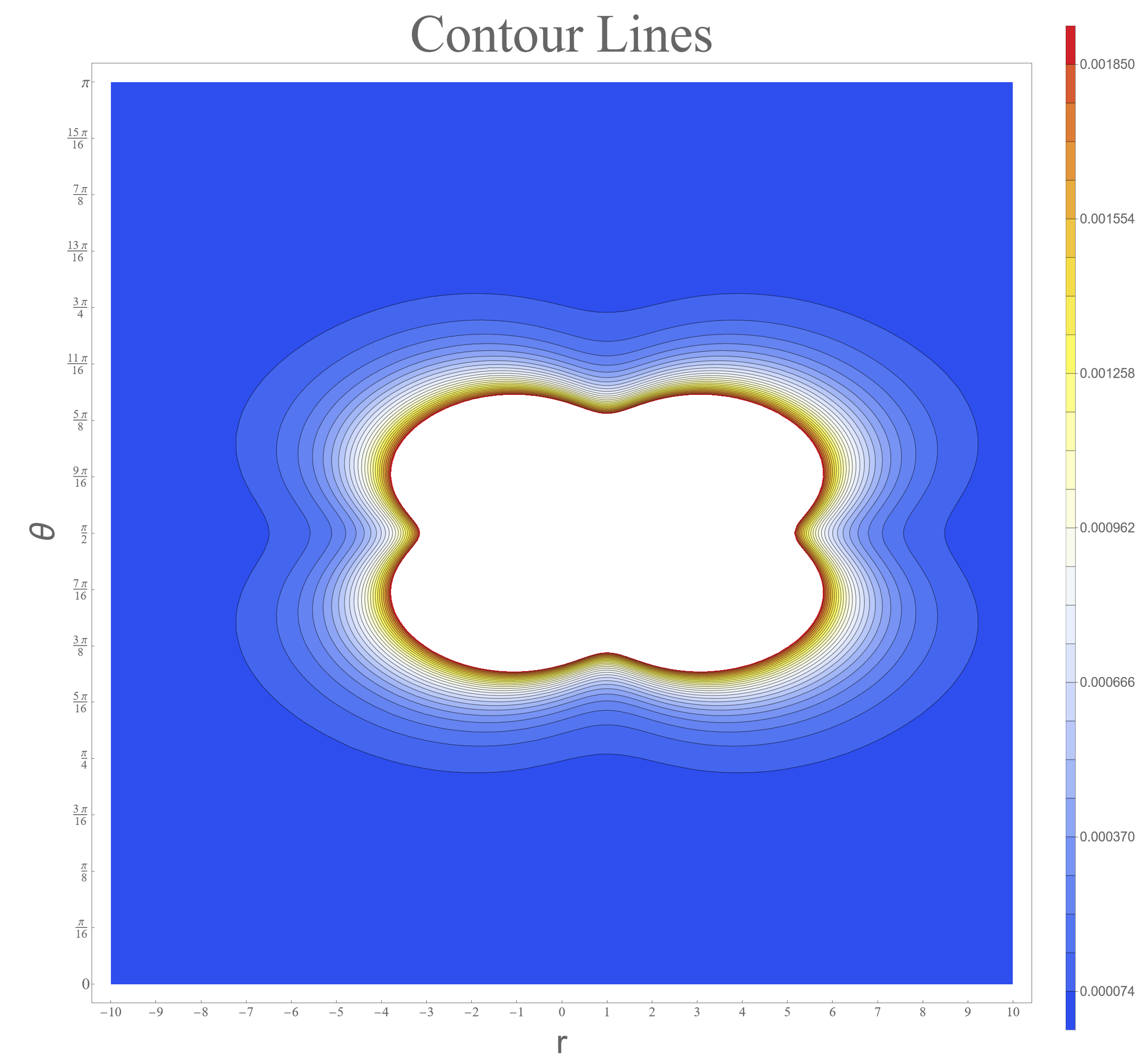}
            \caption{}
            \label{fig:L5_Dilaton-CurvNivel-RDDDD4242}
        \end{subfigure}
    \end{minipage}
    \caption{ 
    Graphs of the cross elements $\tensor{R}{_{ \hat{\mu} \hat{r} \hat{\mu} \hat{r} }} , \quad \hat{\mu}=\hat{\theta},\hat{\varphi}$ associated with tidal forces \textit{for relativistic speed} of $\lambda_5$. The parameters $\lambda_0=1/10^2$, $k_0=1/12$, $l_1=1$, and $L=10$ were used, and the coordinates ($r,\theta$) were selected to enhance the physical visualization. In figure \textbf{a)}, the consequences of choosing either a \textit{phantom or dilatonic scalar field} are compared, using $k_0=\{ -7/12, 1/12 \}$ consecutively and corresponding to the colors \textit{yellow and red.}
    }
    \label{fig:RsCruzadasD_5}
\end{figure*}

Figure [\ref{fig:L5_DilatonFantasma-3D Comparacion-RDDDD2121}] presents a comparative analysis between selecting a phantom or a dilatonic scalar field, where only component $\tensor{R}{_{ \hat{\varphi} \hat{r} \hat{\varphi} \hat{r} }}$ exhibits variation based on the choice of scalar field, while all other components remain unchanged. Likewise, Figures [\ref{fig:L5_Dilaton-3D-RDDDD4242}] and [\ref{fig:L5_Fantasma-3D-RDDDD4242}] pertain to component $\tensor{R}{_{ \hat{\varphi} \hat{r} \hat{\varphi} \hat{r} }}$ under the selection of either a phantom scalar field ($k(x,y>0$) or a dilatonic field ($k(x,y<0$), and their importance is underscored when considering relativistic velocities; Figure [\ref{fig:L5_Dilaton-3D-RDDDD3232}] also holds significance in this context.

Analogous to the component in $\tensor{R}{_{ \hat{\varphi} \hat{t} \hat{\varphi} \hat{t} }}$, the graphs in [\ref{fig:RsCruzadasD_5}] indicate that it is not feasible to access the center of the wormhole without succumbing to tidal forces, particularly when relativistic velocities are considered.

Finally, considering all components and their respective magnitudes, that is, the orders of magnitude among them, it is possible to conclude that the safest angles of entry for both relativistic and non-relativistic velocities are approximately $\theta \approx 0, \pi/2,\pi$, and it is possible to navigate the wormhole without necessitating any alterations in the parameter $\theta$. Thus, the safest entry points for this wormhole lie near the poles, both north and south, and the equatorial plane.

\subsection{Riemann components in \texorpdfstring{$\hat{\mathbb{O}}$}{O-hat} corresponding to \texorpdfstring{$\lambda_{N1}$}{lambda-N1}}

For the case of $\lambda_{N1}$, the components of the Riemann tensor are given by the following expressions:

\begin{subequations}\label{Riemman en SR del astronauta - LambdaN1}

\begin{equation}\label{R2121-LambdaN1}
    \tensor{R}{_{ \hat{x} \hat{t} \hat{x} \hat{t} }}=\frac{e^{-2k(x,y)}}{4L^2 (x^2+y^2)} \big( \lambda_0 x \big)^2 (1-y^2),
\end{equation}
\begin{equation}\label{R3131-LambdaN1}
    \tensor{R}{_{ \hat{y} \hat{t} \hat{y} \hat{t} }} =\frac{e^{-2k(x,y)}}{4L^2 (x^2+y^2) } \big( \lambda_0 y \big)^2 (x^2+1)  ,
\end{equation}
\begin{equation}\label{R4141-LambdaN1}
    \tensor{R}{_{ \hat{\varphi} \hat{t} \hat{\varphi} \hat{t} }} =\frac{e^{-2k(x,y)}}{4L^2 } \lambda_0^2 ,
\end{equation}
\begin{equation}\label{R3232-LambdaN1}
    \tensor{R}{_{ \hat{y} \hat{x} \hat{y} \hat{x} }} = 4 k_0 \tensor{R}{_{ \hat{\varphi} \hat{t} \hat{\varphi} \hat{t} }},
\end{equation}
\begin{multline}\label{R4242-LambdaN1}
    \tensor{R}{_{ \hat{\varphi} \hat{x} \hat{\varphi} \hat{x} }} =  \frac{e^{-2k(x,y)}}{4L^2 (x^2+y^2)} \lambda_0^2 \Bigg\{ 4k_0 y^2 \\ +x^2 \bigg( 3(1-y^2) +4k_0 [2y^2-1] \bigg) \Bigg\}.
\end{multline}
\end{subequations}

\begin{figure*}
\centering
    \begin{minipage}{0.27\textheight}
    \centering    
        \begin{subfigure}{\textwidth}
            \includegraphics[width=\textwidth]{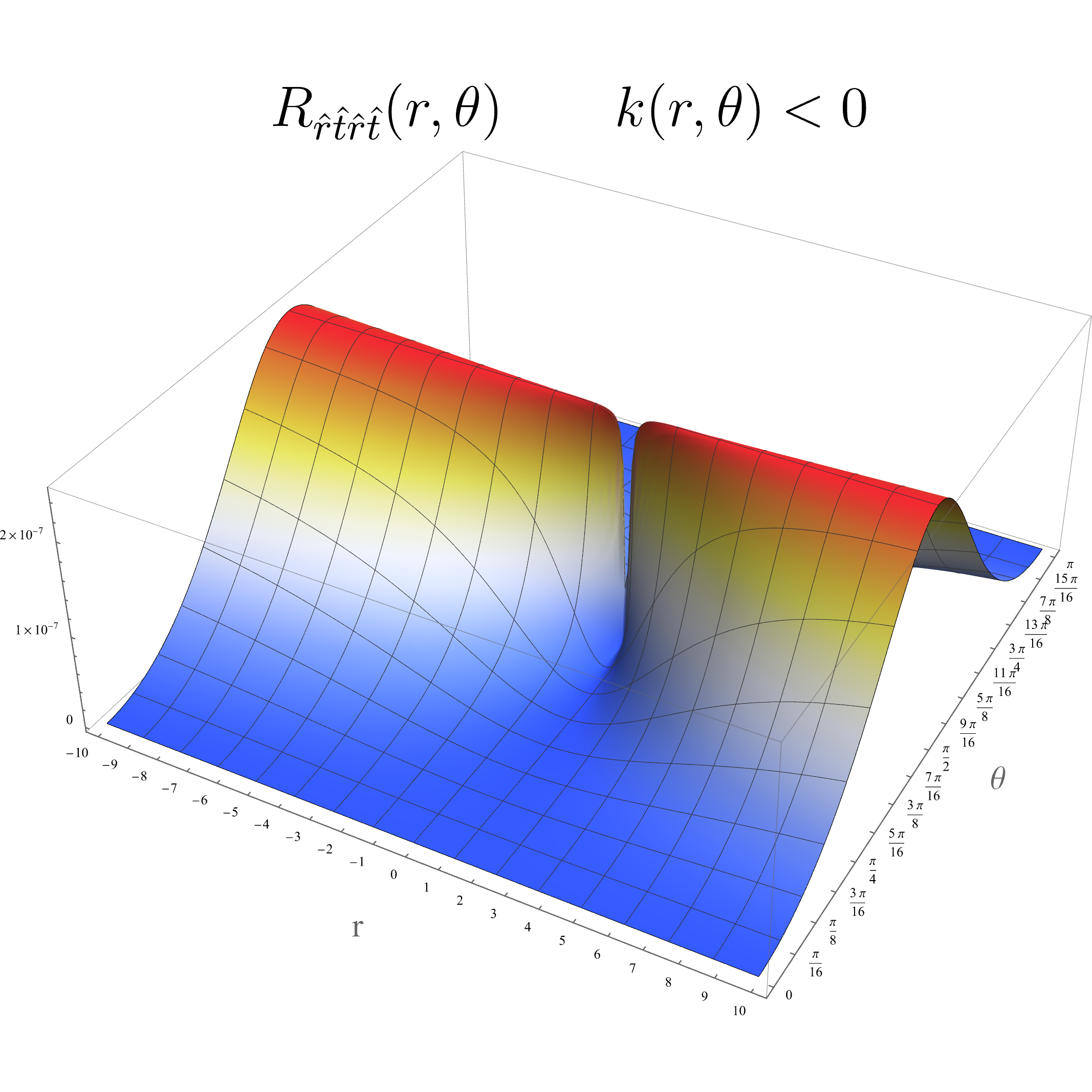}
            \caption{}
            \label{fig:LN1_Dilaton-3D-RDDDD2121}
        \end{subfigure}
    \end{minipage}%
    \hfill
    \begin{minipage}{0.27\textheight}
    \centering    
        \begin{subfigure}{\textwidth}
            \includegraphics[width=\textwidth]{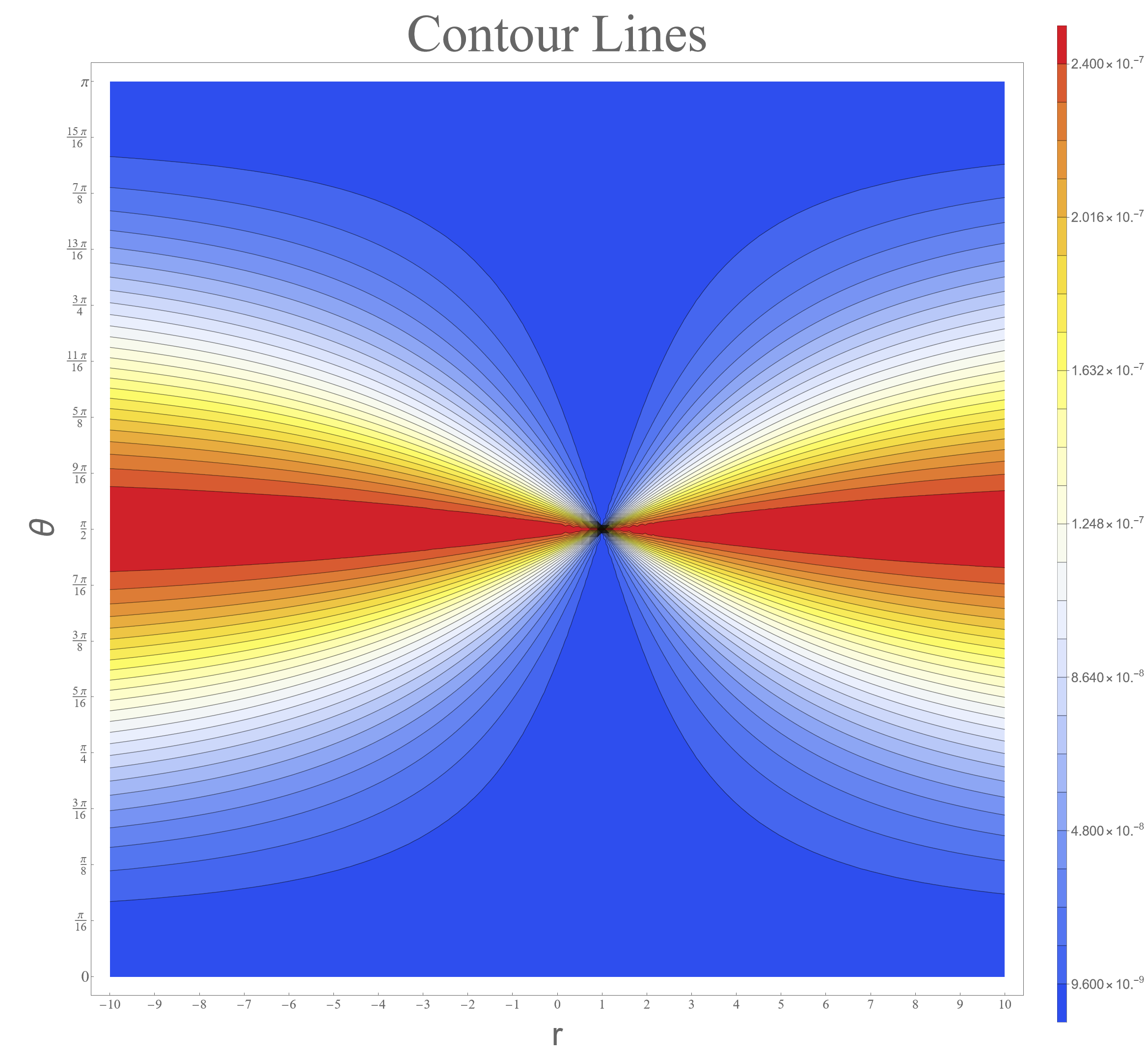}
            \caption{}
            \label{fig:LN1_Dilaton-CurvNivel-RDDDD2121}
        \end{subfigure}
    \end{minipage}
    \centering
    \begin{minipage}{0.27\textheight}
        
        \begin{subfigure}{\textwidth}
            \includegraphics[width=\textwidth]{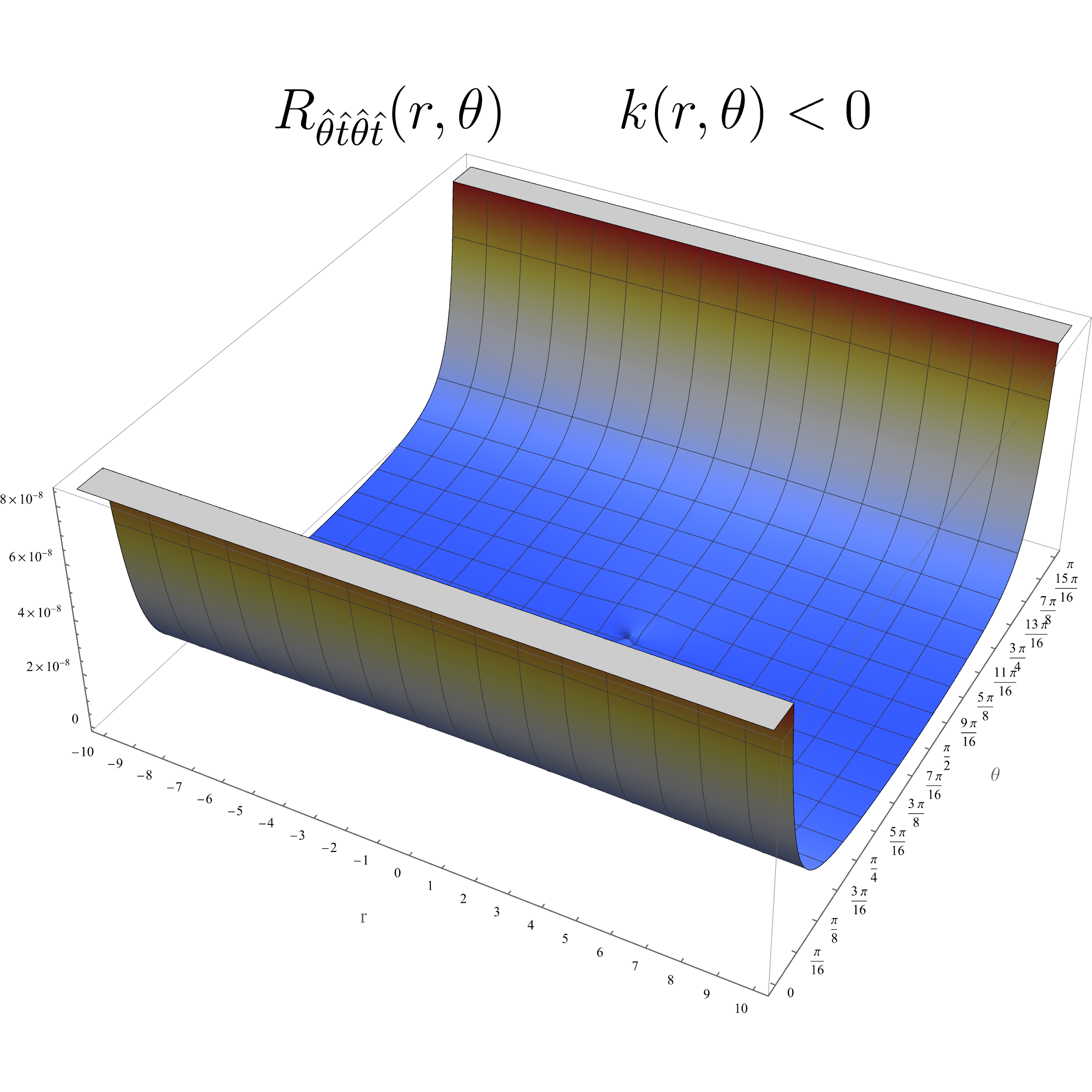}
            \caption{}
            \label{fig:LN1_Dilaton-3D-RDDDD3131}
        \end{subfigure}
    \end{minipage}%
    \hfill
    \begin{minipage}{0.27\textheight}
    \centering    
        \begin{subfigure}{\textwidth}
            \includegraphics[width=\textwidth]{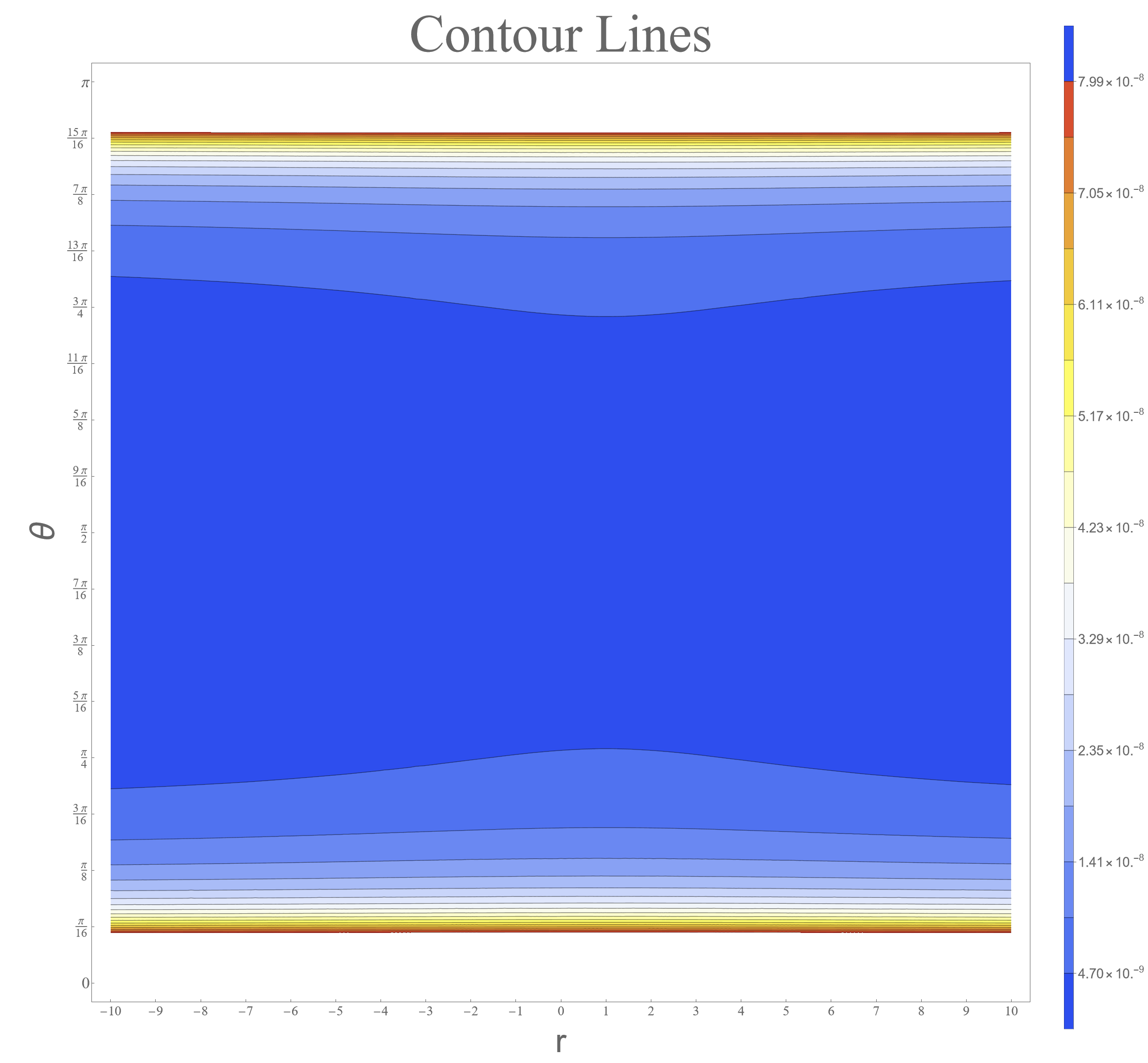}
            \caption{}
            \label{fig:LN1_Dilaton-CurvNivel-RDDDD3131}
        \end{subfigure}
    \end{minipage}
    
    \begin{minipage}{0.27\textheight}
    \centering    
        \begin{subfigure}{\textwidth}
            \includegraphics[width=\textwidth]{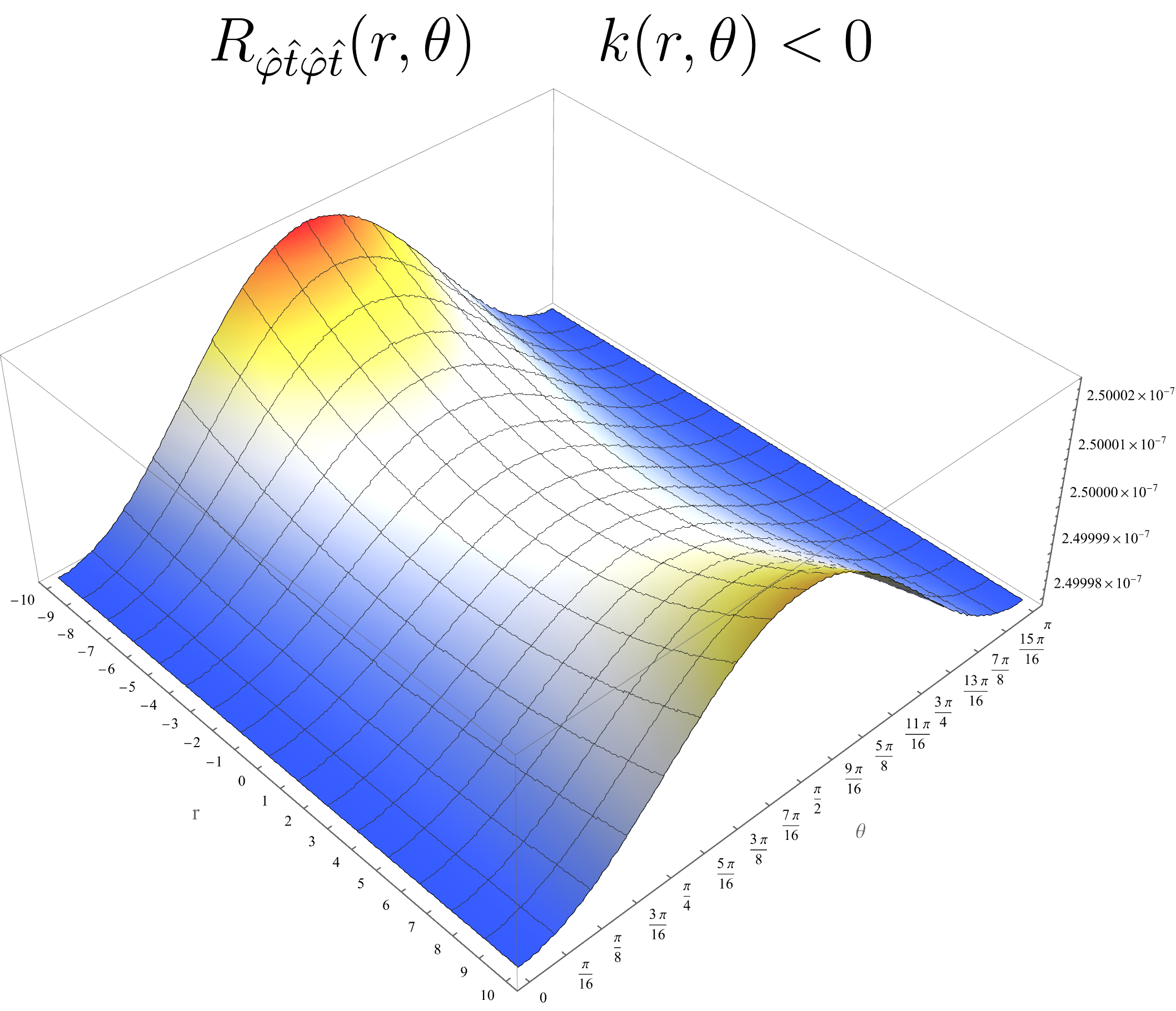}
            \caption{}
            \label{fig:LN1_Dilaton-3D-RDDDD4141}
        \end{subfigure}
    \end{minipage}%
    \hfill
    \begin{minipage}{0.27\textheight}
    \centering
        \begin{subfigure}{\textwidth}
            \includegraphics[width=\textwidth]{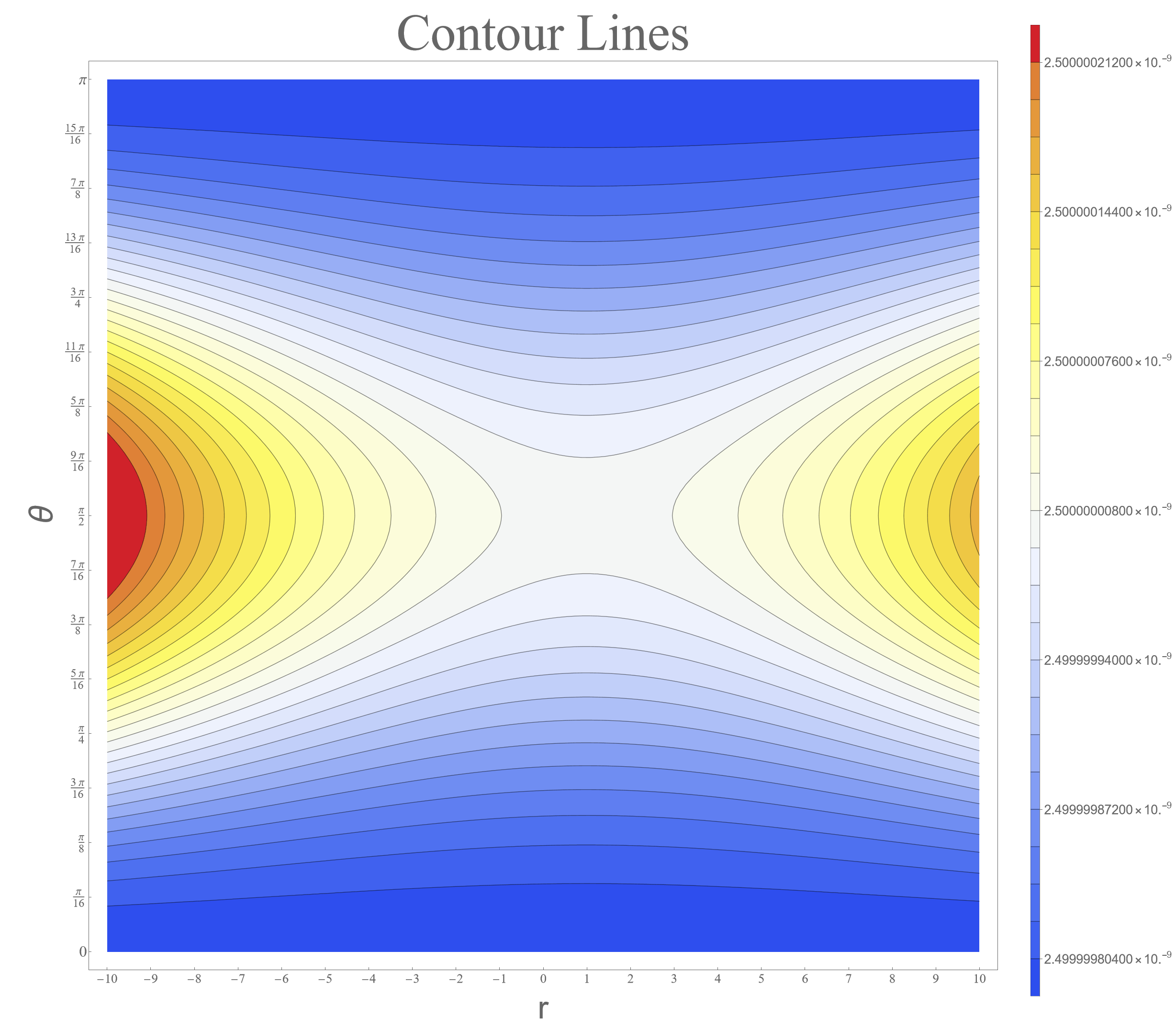}
            \caption{}
            \label{fig:LN1_Dilaton-CurvNivel-RDDDD4141}
        \end{subfigure}
    \end{minipage}
    \caption{ 
    The graphical representations of elements $\tensor{R}{_{ \hat{\mu} \hat{t} \hat{\mu} \hat{t} }} , \quad \hat{\mu}=\hat{r},\hat{\theta},\hat{\varphi}$ associated with $\lambda_5$. The parameters $\lambda_0=1/10^3$, $k_0=1/12$, $l_1=1$, and $L=10$ were used, and the coordinates ($r,\theta$) were selected to enhance the physical visualization. The graphs on the left depict the surface in 3-D generated by $\tensor{R}{_{ \hat{\mu} \hat{t} \hat{\mu} \hat{t} }} , \quad \hat{\mu}=\hat{r},\hat{\theta},\hat{\varphi}$, while the graphs on the right illustrate the contour lines. Both sides employ color schemes relevant to \textbf{high values (red color)} and \textbf{low values (blue color)}. }
    \label{fig:RsNoCruzadasD_N1}
\end{figure*}

\begin{figure*}
    \centering
    \begin{minipage}{0.27\textheight}
    \centering    
        \begin{subfigure}{\textwidth}
            \includegraphics[width=\textwidth]{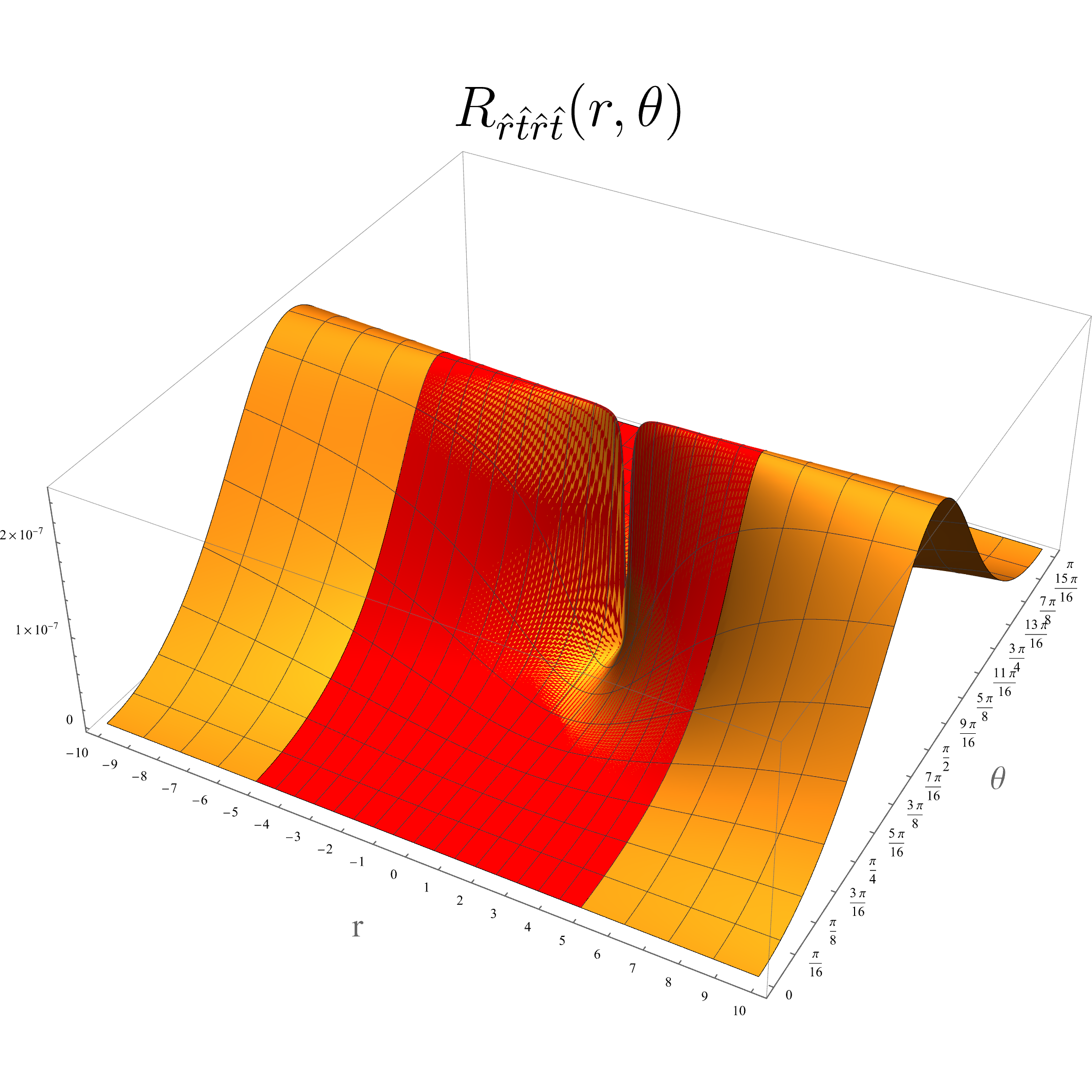}
            \caption{}
            \label{fig:LN1_DilatonFantasma-3D Comparacion-RDDDD2121}
        \end{subfigure}
    \end{minipage}%
    \hfill
    \begin{minipage}{0.27\textheight}
    \centering    
        \begin{subfigure}{\textwidth}
            \includegraphics[width=\textwidth]{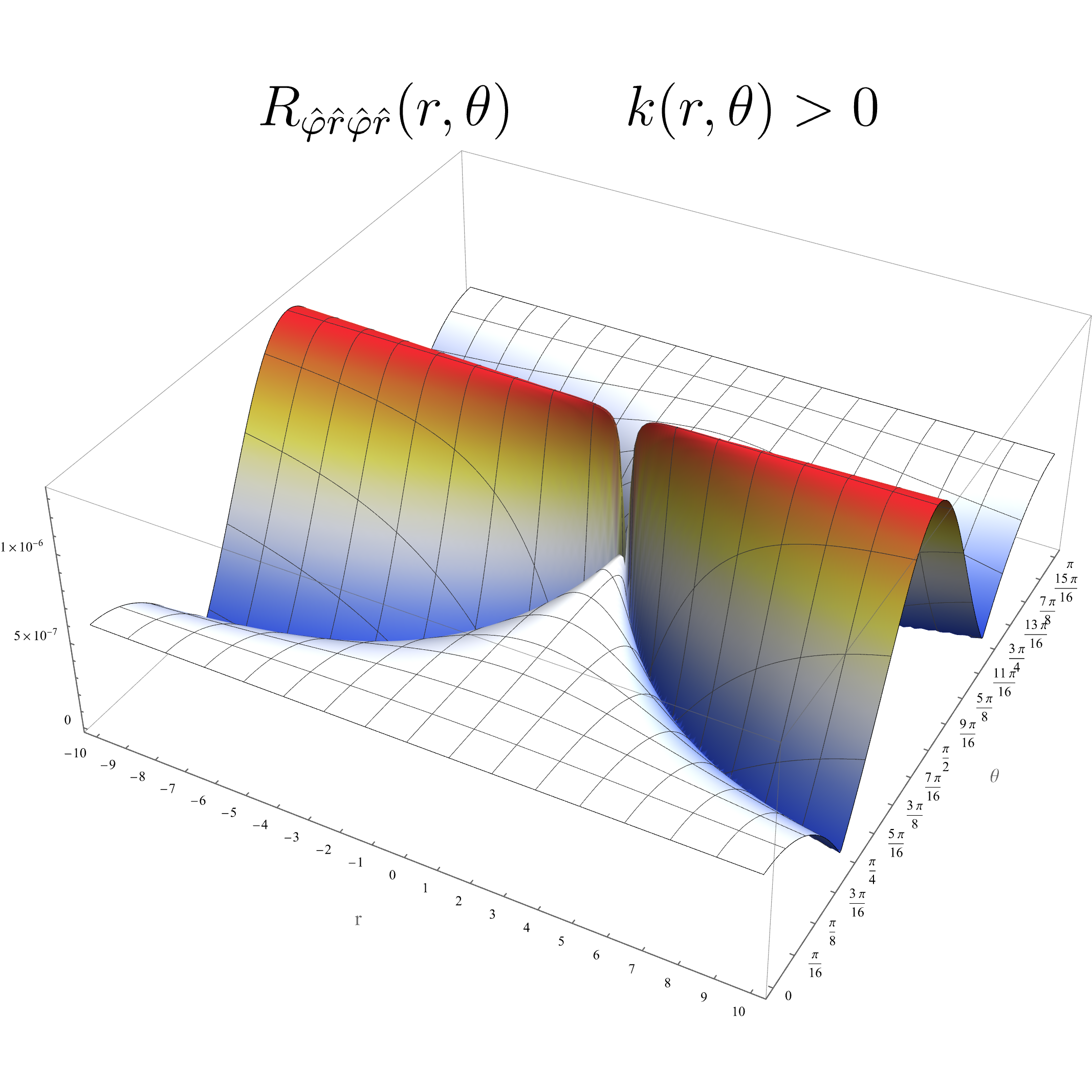}
            \caption{}
            \label{fig:LN1_Fantasma-3D-RDDDD4242}
        \end{subfigure}
    \end{minipage}
    
    \begin{minipage}{0.27\textheight}
    \centering    
        \begin{subfigure}{\textwidth}
            \includegraphics[width=\textwidth]{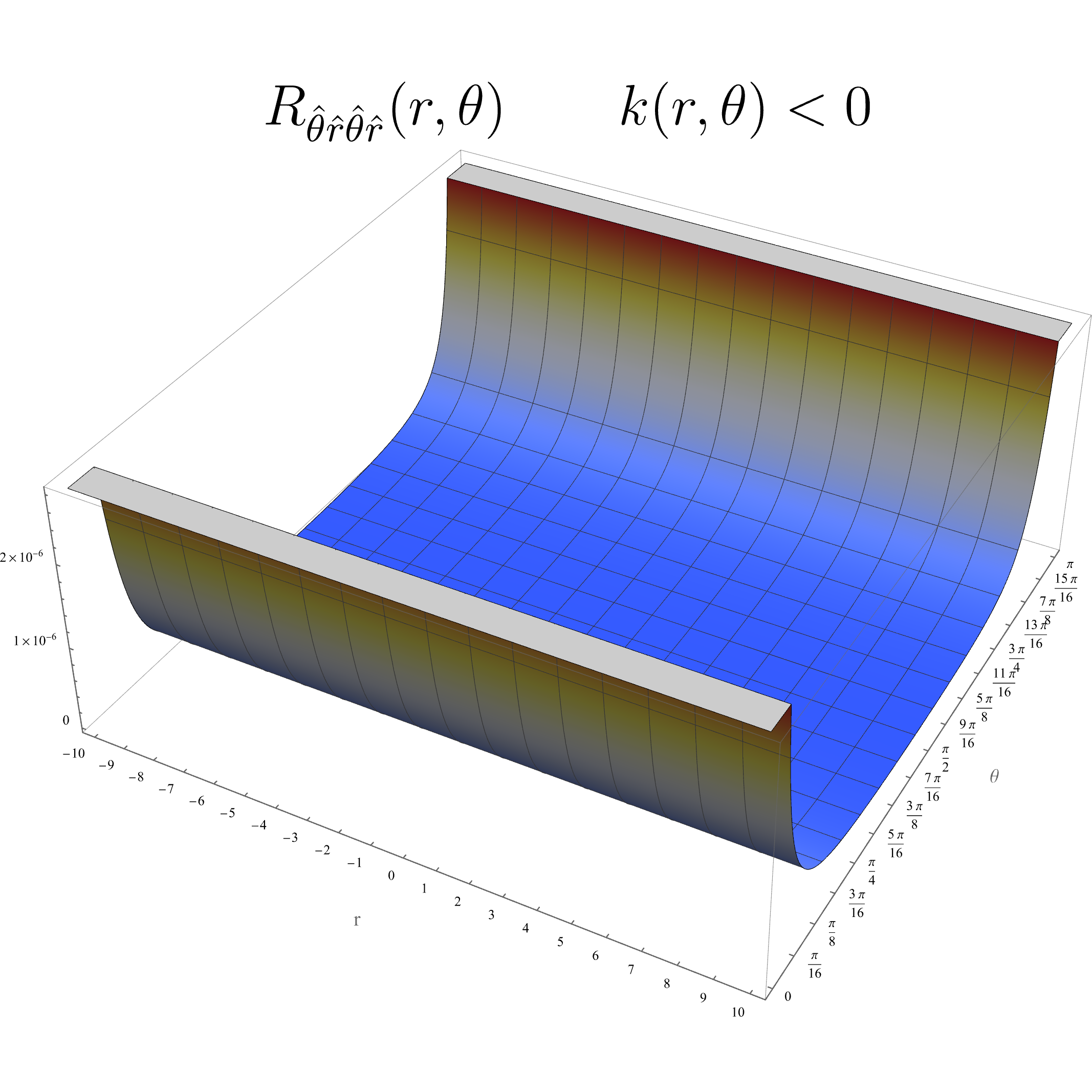}
            \caption{}
            \label{fig:LN1_Dilaton-3D-RDDDD3232}
        \end{subfigure}
    \end{minipage}%
    \hfill
    \begin{minipage}{0.27\textheight}
    \centering    
        \begin{subfigure}{\textwidth}
            \includegraphics[width=\textwidth]{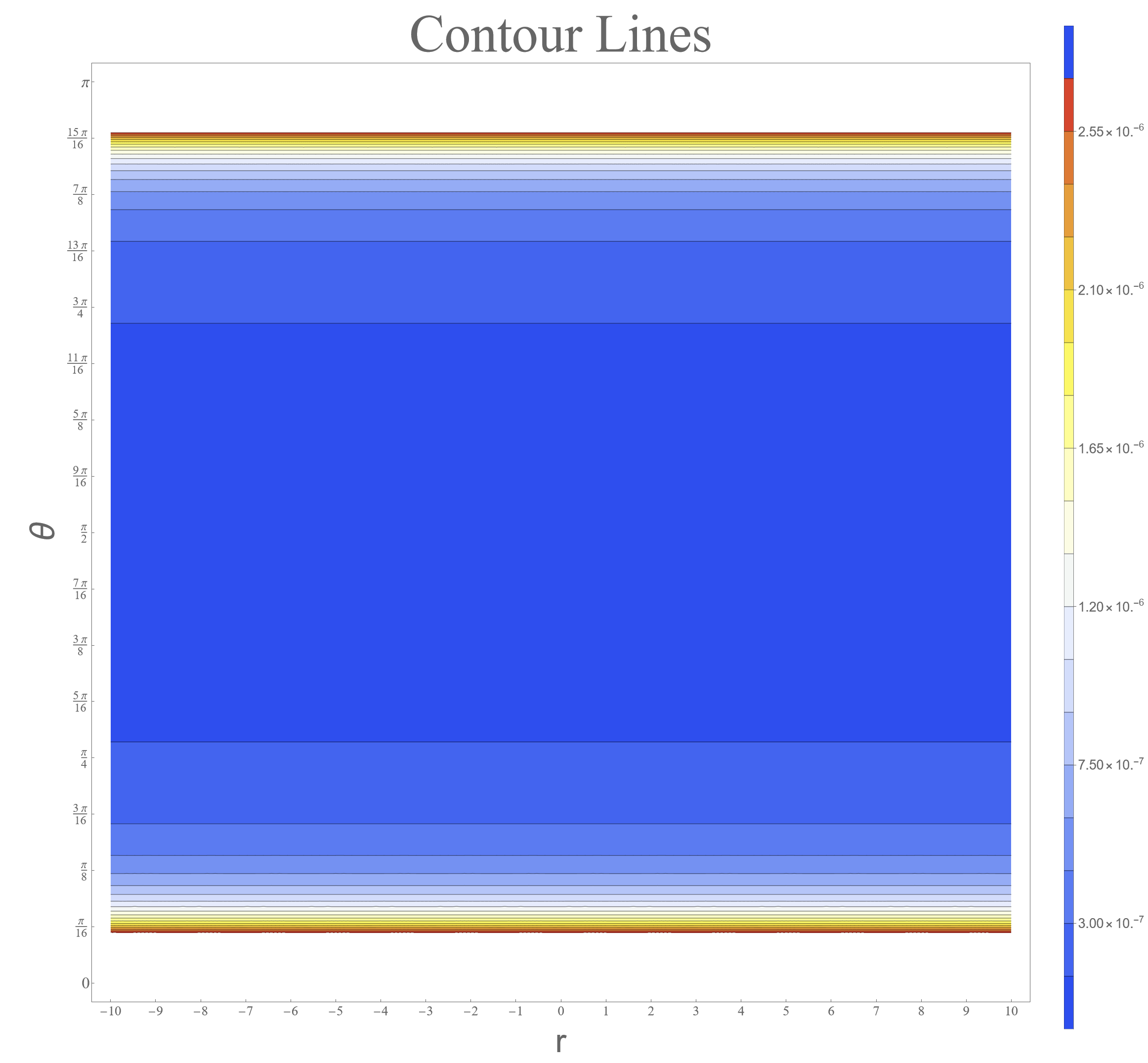}
            \caption{}
            \label{fig:LN1_Dilaton-CurvNivel-RDDDD3232}
        \end{subfigure}
    \end{minipage}
    
    \begin{minipage}{0.27\textheight}
    \centering    
        \begin{subfigure}{\textwidth}
            \includegraphics[width=\textwidth]{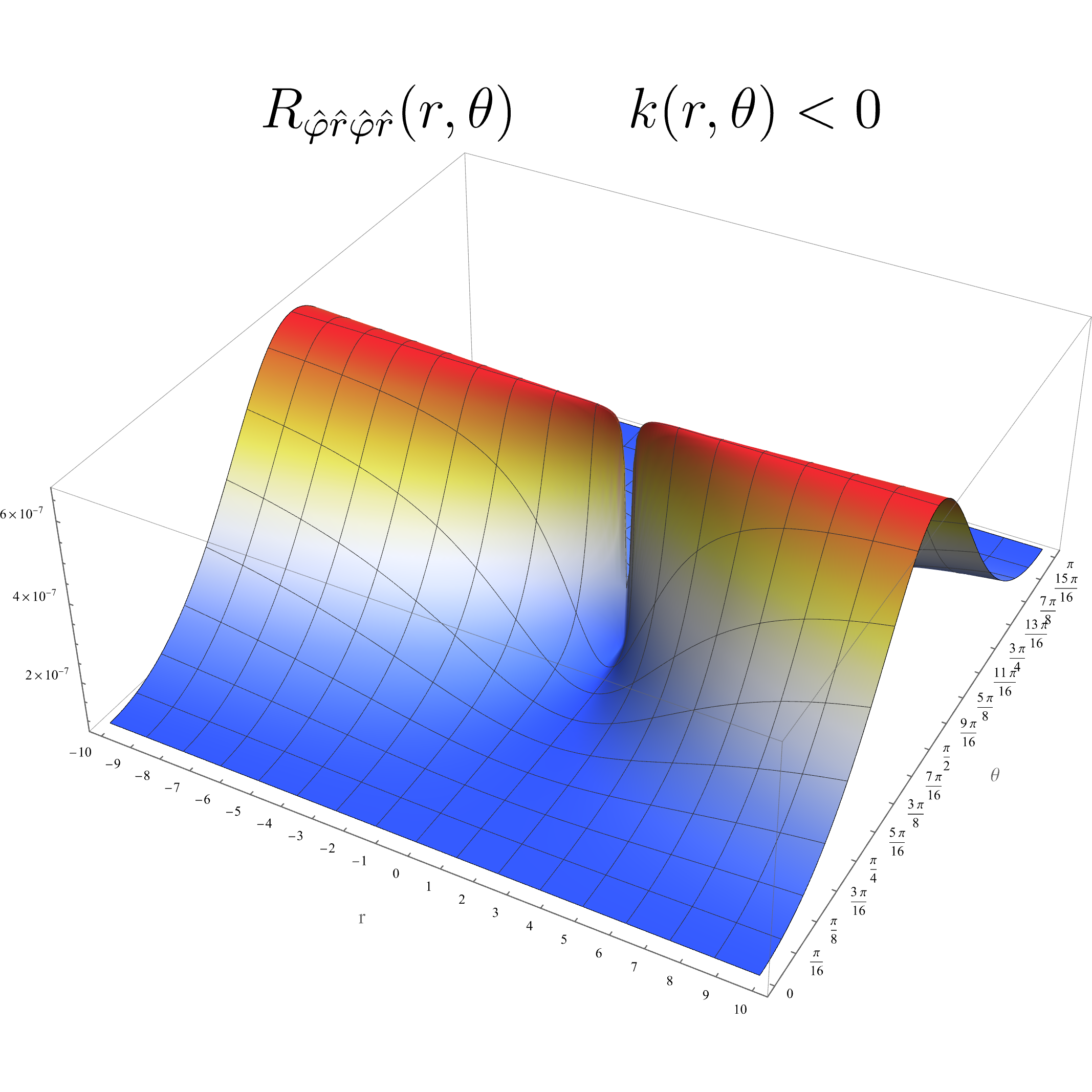}
            \caption{}
            \label{fig:LN1_Dilaton-3D-RDDDD4242}
        \end{subfigure}
    \end{minipage}%
    \hfill
    \begin{minipage}{0.27\textheight}
    \centering    
        \begin{subfigure}{\textwidth}
            \includegraphics[width=\textwidth]{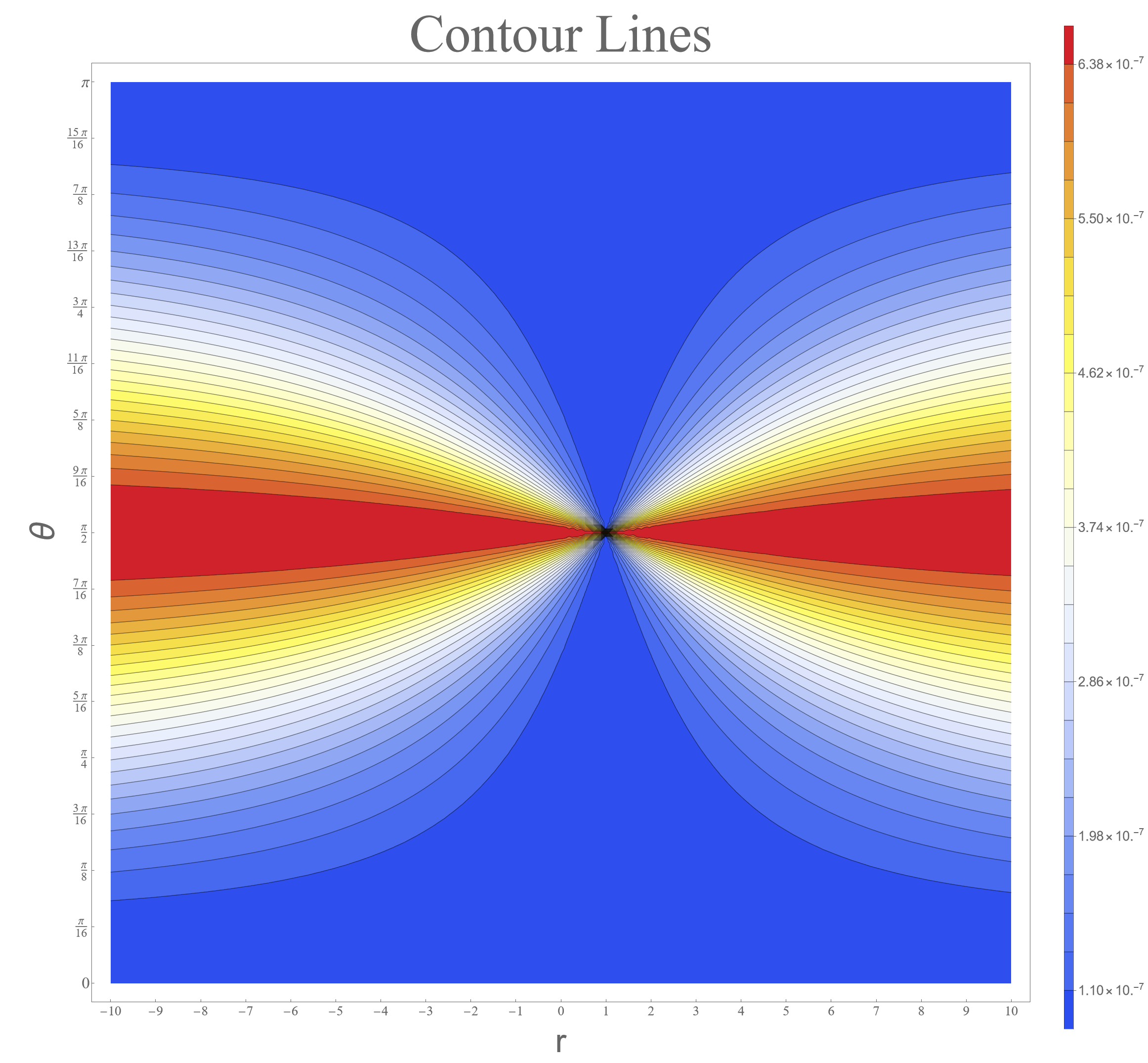}
            \caption{}
            \label{fig:LN1_Dilaton-CurvNivel-RDDDD4242}
        \end{subfigure}
    \end{minipage}
    \caption{ 
    Graphs of the cross elements $\tensor{R}{_{ \hat{\mu} \hat{r} \hat{\mu} \hat{r} }} , \quad \hat{\mu}=\hat{\theta},\hat{\varphi}$ associated with tidal forces \textit{for relativistic speed} of $\lambda_5$. The parameters $\lambda_0=1/10^3$, $k_0=1/12$, $l_1=1$, and $L=10$ were used, and the coordinates ($r,\theta$) were selected to enhance the physical visualization. In figure \textbf{a)}, the consequences of choosing either a \textit{phantom or dilatonic scalar field} are compared, using $k_0=\{ -7/12, 1/12 \}$ consecutively and corresponding to the colors \textit{yellow and red.}
    }
    \label{fig:RsCruzadasD_N1}
\end{figure*}

Figures [\ref{fig:RsNoCruzadasD_N1}] and [\ref{fig:RsCruzadasD_N1}] display all the graphs of the elements. Figure [\ref{fig:RsNoCruzadasD_N1}] shows the key elements in the case of non-relativistic velocities, once again for a dilatonic scalar field ($k_0=1/12$). Figure [\ref{fig:RsCruzadasD_N1}] presents the relevant components for the relativistic velocity case. In Figure [\ref{fig:LN1_DilatonFantasma-3D Comparacion-RDDDD2121}], it is shown once again that there are no significant differences in the choice of the scalar field type, except for the component  $\tensor{R}{_{ \hat{\varphi} \hat{r} \hat{\varphi} \hat{r} }}$ , as seen in Figures [\ref{fig:LN1_Dilaton-3D-RDDDD4242}] and [\ref{fig:LN1_Fantasma-3D-RDDDD4242}].

Using the graphs to better analyse, we can observe that the behaviour of ($\lambda_{N1}$) is very similar to the corresponding ($\lambda_5$) solution. In this case, the optimal entry angles to avoid destruction are not limited to regions near the poles, as the components $\tensor{R}{_{ \hat{r} \hat{t} \hat{r} \hat{t} }}$, $\tensor{R}{_{ \hat{\varphi} \hat{t} \hat{\varphi} \hat{t} }}$ and $\tensor{R}{_{ \hat{\varphi} \hat{t} \hat{\varphi} \hat{t} }}$ are bounded along the z-axis, allowing entry from various angles. However, if we wish to completely avoid any tidal forces, it is necessary to enter regions near the poles; but, in this case, we will need to vary $\theta$.

An interesting point is that when relativistic velocities and a phantom scalar field are considered, in order to completely avoid tidal forces, it is necessary to follow the path defined by Figure [\ref{fig:LN1_Fantasma-3D-RDDDD4242}].
\section{Geodesics}

To obtain the geodesic curves, we used the following equation: 

\begin{equation}\label{EcuacionDeLaGeodesica}
    \frac{d^2 x^\mu}{d \tau^2} = -\Gamma^\mu_{\alpha \beta} \frac{dx^\alpha}{d\tau} \frac{dx^\beta}{d\tau}.
\end{equation}

We employ numerical methods to solve equation (\ref{EcuacionDeLaGeodesica}), using the same parameters as specified in Section \textit{Tidal Forces}. The initial conditions for all graphs were $t(0)=0$, $x(0)=25$, $y(0)=\{0.01,0.25,0.50,0.75,0.95 \}$, $\varphi(0)=0$, $t'(0)=1$, $x'(0)=-1$, $y'(0)=0$, and $\varphi '(0)=0$. The program used was \textit{Wolfram Research, Inc., Mathematica, Version 14.0.0.0, Champaign, IL (2024)}, and the numerical method used was \textit{Herminte, 3-degree,}\footnote{The Hermite method is a high-order interpolation method suitable for smooth solutions, automatically selected in NDSolve when the equations and integration domain meet specific smoothness and stability conditions.} with a precision goal and an accuracy goal of 10 digits.

\begin{figure}
    %\centering
    \begin{minipage}{0.3\textheight}
        \includegraphics[width=\textwidth]{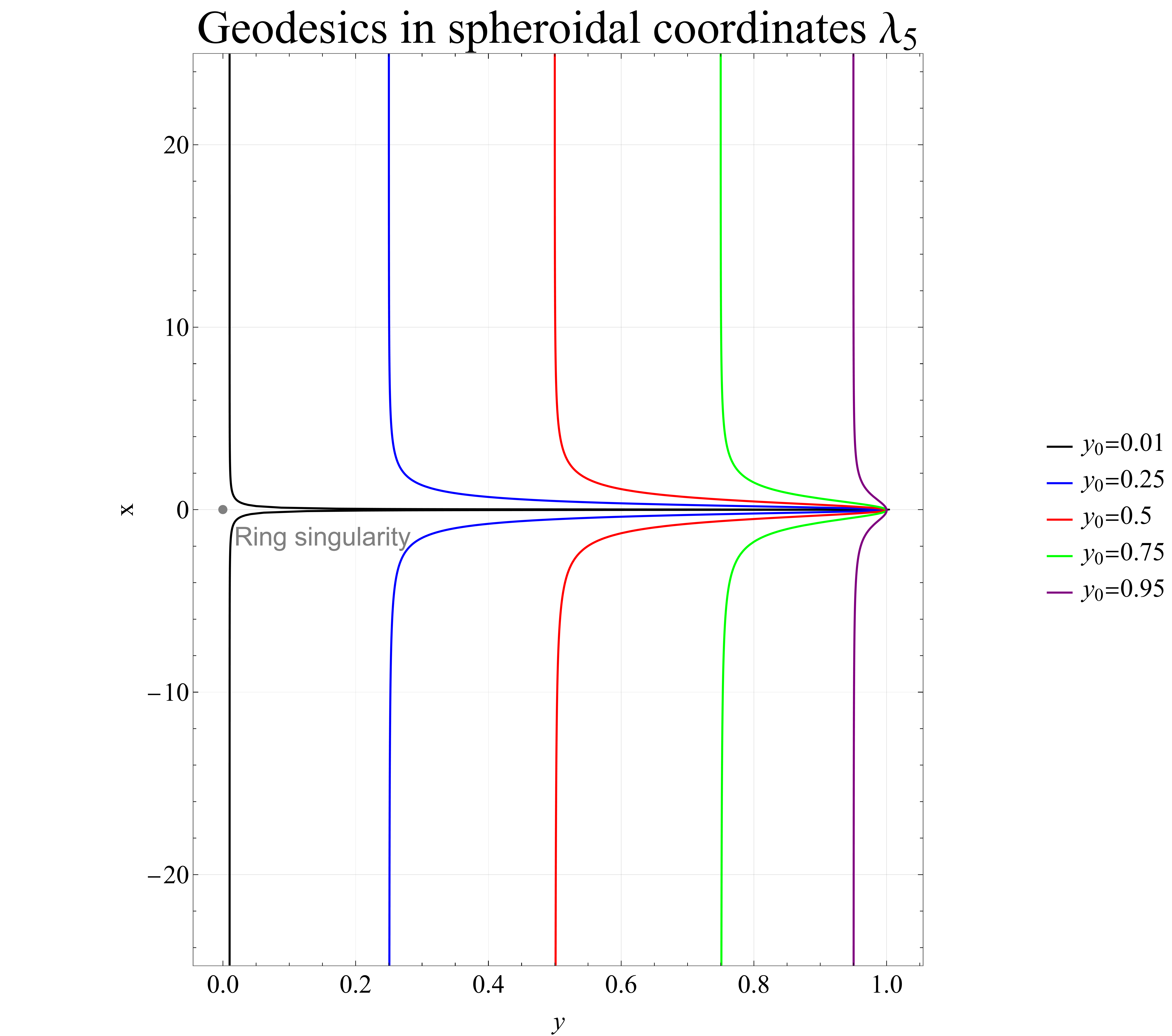}
        \subcaption{Graphs pertinent to the geodesics of the $\lambda_{5}$ solution in spheroidal coordinates are presented. Diverse values of $y_0$ were used to derive the curves, along with the initial conditions: $t(0)=0$, $x(0)=25$, $y(0)=\{0.01,0.25,0.50,0.75,0.95 \}$, $\varphi(0)=0$, $t'(0)=1$, $x'(0)=-1$, $y'(0)=0$, $\varphi '(0)=0$, as well as the parameters $f=1$, $L=10$, $\lambda_0=1/10^2$. }
        \label{fig:GeodesicasL5}
    \end{minipage}
    \hfill
    \begin{minipage}{0.3\textheight}
        \includegraphics[width=\textwidth]{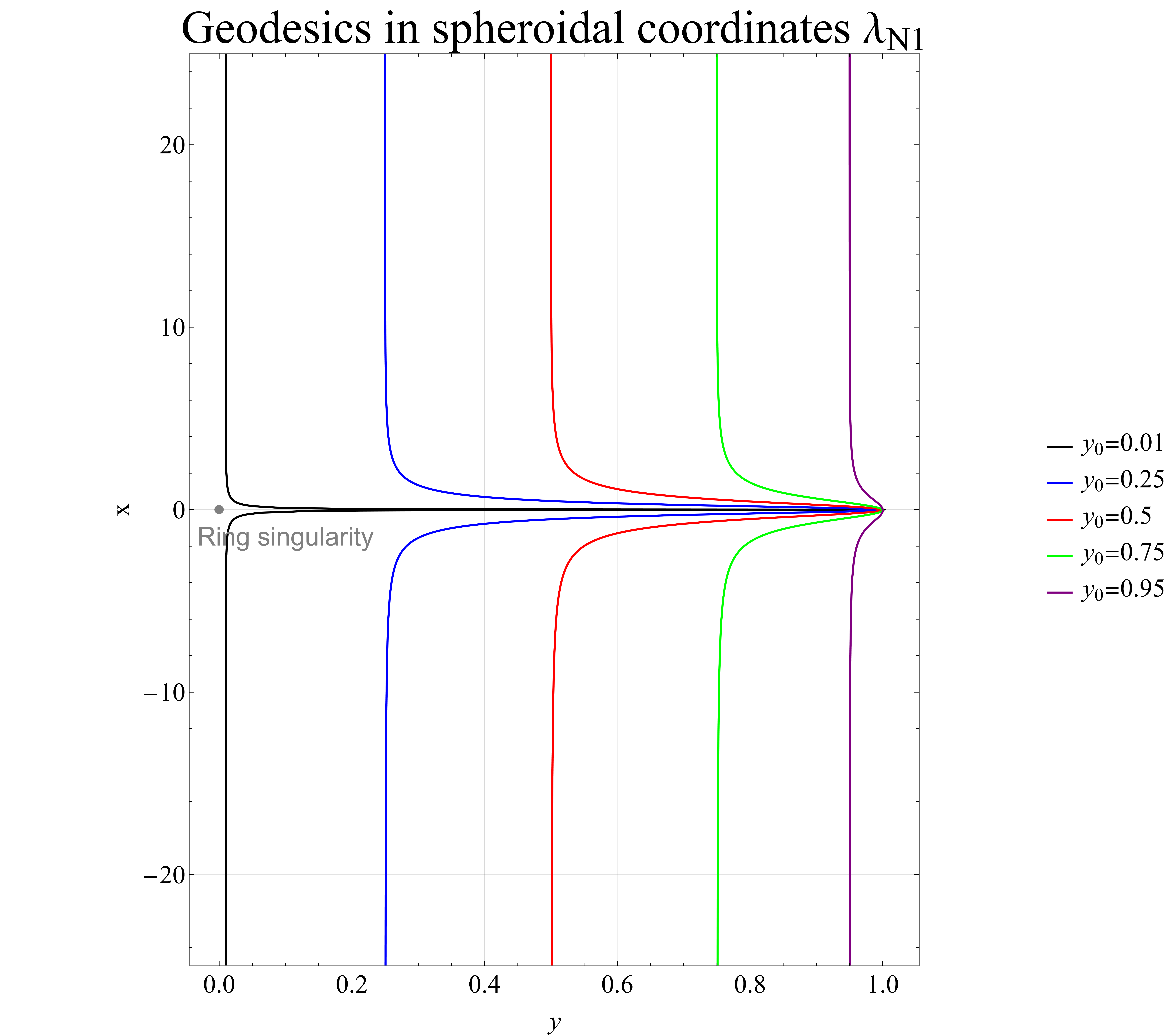}
        \subcaption{Graphs pertinent to the geodesics of the $\lambda_{N1}$ solution in spheroidal coordinates are presented. Diverse values of $y_0$ were used to derive the curves, along with the initial conditions: $t(0)=0$, $x(0)=25$, $y(0)=\{0.01,0.25,0.50,0.75,0.95 \}$, $\varphi(0)=0$, $t'(0)=1$, $x'(0)=-1$, $y'(0)=0$, $\varphi '(0)=0$, as well as the parameters $f=1$, $L=10$, $\lambda_0=1/10^3$.  }
        \label{fig:GeodesicasLN1}
    \end{minipage}
    \caption{\label{fig:GeodesicasAmbas}The values for $x>0$ pertain to one universe, whereas those for $x<0$ correspond either to an alternate universe or to the same universe at a distinct location for both graphs. It should be noted that the geodesic associated with $y_0=0$ in both figures, approaches the ring singularity without making contact; the geometry provokes a rebound, causing it to return with an exceptionally high velocity. }
\end{figure}

For both solutions, the geodesics were plotted in spheroidal coordinates ($x,y$) to better interpret the behavior of the curves. Figure [\ref{fig:GeodesicasL5}] corresponds to $\lambda_{5}$, while Figure [\ref{fig:GeodesicasLN1}] corresponds to $\lambda_{N1}$. For both scenarios, geodesic $y_0=0$ is excluded due to extreme sensitivity near the ring singularity, which poses significant challenges for numerical plotting.

In the solutions presented, the geodesics are observed to navigate the wormhole smoothly. Most notably, when a geodesic is directed toward the ring singularity, the compact structure of the spacetime deflects it, either redirecting the trajectory to circumvent the singularity or causing it to return to its original universe. This dynamic inherently prevents any direct encounter with the ring singularity, ensuring a smooth passage through the wormhole.

\section{Units}

The key to interpreting the physics of these objects lies in the units. To determine the magnitudes of the electromagnetic fields and rotation, it is necessary to know the units of all the variables used.

We start with the Lagrangian, which has energy density units in the International System of Units (SI):

\begin{multline}\label{LagrangianoTesisUnidades}
    \mathfrak{L}=\sqrt{-g}\bigg(-\frac{c^4}{8 \pi G}R +\frac{c^4}{8 \pi G}2\epsilon_0 (\nabla \phi)^2 \\ + \frac{1}{\mu_0}e^{-2 \alpha_0 \phi } F^2 \bigg),
\end{multline}
where $c$ is the speed of light, $G$ is the gravitational constant and, for this occasion we will use $\varepsilon_0$ as vacuum permittivity and $\mu_0$ vacuum permeability. And for:
\begin{align*}
    R \quad &[=] \quad (\text{Length})^{-2},\\
    \{\phi,\alpha_0,\epsilon_0\} \quad &[=] \quad \text{Dimensionless}, \\
    F_{\mu \nu} \quad &[=] \quad \text{Teslas}.
\end{align*}

The square line element in cylindrical coordinates $ds$ takes the following form:
\begin{multline}\label{ds Cilindricas}
    ds^2=-f\left\{ d(ct)-\frac{\omega}{L} d (L\varphi) \right\}^2 \\+f^{-1} \left\{ e^{2k} (d\rho ^2 + dz^2) +\frac{\rho^2}{L^2} d(L\varphi)^2 \right\},
\end{multline}

where $f$ and $k$ are dimensionless, and $\{\rho,z,\omega, ct\}$ has length units.  When spheroidal coordinates are considered (\ref{ds sp}), 

\begin{align*}
    \{x,y \} \quad &[=] \quad \text{Dimensionless},\\
    L \quad &[=] \quad \text{Length},
\end{align*}
 with $L$ providing the units to the coordinates.

To obtain the correct units for the four-potential, we will use:
\begin{equation}\label{ConstanteL}
    \sigma_0 \equiv \frac{8\pi G}{\mu_0 c^4} \qquad \Rightarrow \quad \frac{1}{\sqrt{\sigma_0}}\thickapprox 2.46 \times 10^{18}   [\text{Meters*Teslas}].
\end{equation}
Then
\begin{multline}
    A_{\mu} = \big[A_t,0,0,A_{\varphi}\big]= \big[\Phi/c,0,0,A_{\varphi}\big] \\=\frac{\sqrt{f_0}}{2 \kappa_0 \sqrt{\sigma_0} }e^{-\lambda}\left[ 1,0,0,- \frac{\omega}{L} \right],
\end{multline}
where $\Phi$ is the electric potential, and has volt unit and $\lambda$ is dimensionless. 

Therefore, the electromagnetic field, takes the form:

\begin{widetext}
\begin{equation}\label{CampoMagnetico}
    \begin{bmatrix} B_z \\ -B_\rho \end{bmatrix}=\frac{-1/\sqrt{\sigma_0}}{L(x^2+y^2)}\frac{\sqrt{f_0}}{2\kappa_0}
    \begin{bmatrix}
        x \sqrt{(x^2+1)(1-y^2)} & -y \sqrt{(x^2+1)(1-y^2)} \\
        y(x^2+1) & x(1-y^2)
    \end{bmatrix}
    \begin{bmatrix} \partial_{x} \\\partial_{y} \end{bmatrix} \left( \frac{\omega}{L} e^{-\lambda} \right),
\end{equation}
\begin{equation}\label{CampoElectrico}
    \begin{bmatrix} E_\rho \\ E_z \end{bmatrix}=\frac{c/\sqrt{\sigma_0}}{L(x^2+y^2)}\frac{\sqrt{f_0}}{2\kappa_0}
    \begin{bmatrix}
        x \sqrt{(x^2+1)(1-y^2)} & -y \sqrt{(x^2+1)(1-y^2)} \\
        y(x^2+1) & x(1-y^2)
    \end{bmatrix}
    \begin{bmatrix} \partial_{x} \\\partial_{y} \end{bmatrix} \left(  e^{-\lambda} \right),
\end{equation}
\end{widetext}

in which $E$ denotes the electric field and $B$ represents the magnetic field, with respective units of tesla and volt/meter.

If we consider a test particle near the wormhole, we can derive the angular velocity induced by the wormhole on the particle.

Recall that the contravariant and covariant components of the angular momentum of a particle are given by:

\begin{equation*}
    p^{\alpha}=g^{\alpha \beta}p_{\beta}.
\end{equation*}

Taking into account the metric (\ref{ds Cilindricas}), the $t$ and $\phi$ components of the contravariant angular momentum are:

\begin{align*}
    p^{\varphi}&=g^{\varphi \varphi}p_{\varphi}+g^{\varphi t}p_{t},\\
    p^{t}&=g^{t \varphi}p_{\varphi}+g^{t t}p_{t}.
\end{align*}

Now, assuming that the particle is solely influenced by the spacetime deformation caused by the object, we set $p_{t}$ as the only non-zero component, while all other components are zero. Taking these considerations into account, the angular velocity is given by:

\begin{equation}
\dot{\varphi}=\frac{d \varphi}{dt}=\frac{f \omega}{f^2 \omega^2 - L^2(x^2+1)(1-y^2)}c.
\end{equation}

Finally, by drawing an analogy with black holes and considering the physical meaning of the constants accompanying the variable transformations \footnote{There exists a transcendental condition, $L<2l_1$, i.e., the throat is situated within the area defined by the Schwarzschild radius.}
\begin{equation}
    Lx=r-l_1 , \quad z=L x y, \qquad L^2+l_1 ^2=l_0^2,
\end{equation}

we obtain the following definitions.
\begin{align}
    l_1 =\frac{ M G }{c^2} \qquad &[=] \quad\text{Length},\\
    J_k \equiv \frac{2GJ}{c^3} \qquad &[=] \quad\text{Length}^2,\\
    a\equiv \frac{J_k}{2l_1}\qquad &[=] \quad\text{Length} ,\\
    Q_{L}^2 \equiv \frac{G}{c^4}\frac{Q^2}{\varepsilon_0} \qquad &[=] \quad\text{Length}^2 ,\\
    H_{L}^2 \equiv \frac{G}{c^4}\frac{H^2}{\mu_0} \qquad &[=] \quad\text{Length}^2 ,\\ 
    l_0^2 \equiv a^2 +Q_L^2+H_L^2 \qquad &[=] \quad\text{Length}. \label{Relacion WH}
\end{align}

In this context, $R_s=2l_1$ denotes the Schwarzschild radius, $J$ is characterized by having angular momentum as its unit, $M$ is associated with mass as its unit, and $\{ Q, H \}$ are described by the units electric charge and magnetic charge, respectively.

This paper will present three instances of wormholes and detail their respective parameters in two tables. Assuming the wormhole's shape is an oblate spheroid $R_p = R_e/3$, which means that the equatorial radius is three times the dimension of the polar radius, then, the angular momentum formula is:
\begin{equation}\label{Momento Angular kg}
    J = \frac{10}{45} M r^2 \dot{\varphi}.
\end{equation}

The angular velocity can be computed considering $\theta=\pi/4$,  $\lambda_0=1/10^2$ for $\lambda_5$ and $\lambda_0=1/10^3$ for $\lambda_{N1}$. Consequently, this allows for the determination of the variable $a$, thereby allowing the acquisition of $Q_L^2+H_L^2$ for each value of $L < R_s=2l_1$ for $\lambda_{5}$ and for $\lambda_{N1}$. The results will be shown in the \textbf{Table} (\ref{ValoresL5}), and (\ref{ValoresLN1}). \textit{The electromagnetic field related to $\lambda_{N1}$ represents an exception as it exhibits an exponential decrease.}

\begin{table*}
\caption{Dimension of the parameters of the WH ($\lambda_5$). The first row, consisting of three values of $L$, corresponds to a Sun-like wormhole, i.e., one with a mass equal to that of the Sun, $M=M_{\odot} \approx 1.989\times 10^{30}$ kg, consequently, the Schwarzschild radius is $R_s \approx 2.953 \times 10^3$ m. To determine the rotation parameters, a radius equivalent to the Sun's radius, $r \approx7 \times 10^8$ m, is considered.  The second row, composed of three values of $L$, corresponds to a magnetar-type wormhole, i.e., one with $M=2 M_{\odot}$, $R_s \approx 6 \times 10^3$ m, and a radius of $r \approx 10^5$ m.  Finally, the last row, also consisting of three values of $L$, corresponds to a wormhole the size of a supermassive black hole, with parameters $M=10^7\, M_{\odot}$, $R_s \approx 3 \times 10^{10}$ m, and a radius of $r \approx 10^{12}$ m.  }
\label{ValoresL5}
\begin{ruledtabular}
\begin{tabular}{cccccc}
\makecell{$L$ \\ (Meters)} & 
\makecell{$\dot{\varphi}$ \\ (Seconds$^{-1}$)} & 
\makecell{$a$ \\ (Meters)} & 
\makecell{$Q_L^2+H_L^2$ \\ (Meters)} &
\makecell{$E$ \\ (Volt/Meters)} &
\makecell{$B$ \\ (Teslas)} \\
\hline
$ 10^{1}$  &  $8.653 \times 10^{-11}$ & $3.143 \times 10^{-2}$  & $2.181 \times 10^{6}$ & $1.064 \times 10^{8}$  & $-1.757 \times 10^{7}$ \\
$ 10^{2}$  & $8.653 \times 10^{-10}$ & $3.143 \times 10^{-1}$  & $2.191 \times 10^{6}$ & $1.064 \times 10^{9}$  & $-1.757 \times 10^{7}$\\
$ 10^{3}$   & $8.653 \times 10^{-9}$  & $3.143 \times 10^{0}$   & $3.181 \times 10^{6}$ & $1.064 \times 10^{10}$ & $-1.757 \times 10^{7}$\\
\hline
$ 10^{1}$  &  $4.501 \times 10^{-3}$ & $3.337 \times 10^{-2}$  & $8.722 \times 10^{6}$ & $5.537 \times 10^{15}$  & $-1.267 \times 10^{11}$ \\
$ 10^{2}$  & $4.501 \times 10^{-2}$ & $3.337 \times 10^{-1}$  & $8.732 \times 10^{6}$ & $5.537 \times 10^{16}$  & $-1.267 \times 10^{11}$\\
$ 10^{3}$   & $4.501 \times 10^{-1}$  & $3.337 \times 10^{0}$   & $9.722 \times 10^{6}$ & $5.537 \times 10^{17}$ & $-1.267 \times 10^{11}$\\
\hline
$ 10^{1}$  &  $4.368 \times 10^{-17}$ & $3.238 \times 10^{-2}$  & $2.181 \times 10^{20}$ & $5.372 \times 10^{1}$  & $-1.248 \times 10^{4}$ \\
$ 10^{3}$  & $4.368 \times 10^{-15}$ & $3.238 \times 10^{0}$  & $2.181 \times 10^{20}$ & $5.372 \times 10^{3}$  & $-1.248 \times 10^{4}$\\
$ 10^{5}$   & $4.368 \times 10^{-13}$  & $3.238 \times 10^{2}$   & $2.181 \times 10^{20}$ & $5.372 \times 10^{5}$ & $-1.248 \times 10^{4}$\\
\end{tabular}
\end{ruledtabular}
\end{table*}

\begin{table}
\caption{Parameters for a  Sun-like wormhole ($\lambda_{N1}$), with $M=M_{\odot} \approx 1.989\times 10^{30}$ kg, which results in a Schwarzschild radius of $R_s \approx 2.953 \times 10^3$ m. To evaluate the rotational parameters, a radius equivalent to the Sun's, $r \approx 7 \times 10^8$ m, is used.}
\label{ValoresLN1}
\begin{ruledtabular}
\begin{tabular}{cccc}
\makecell{$L$ \\ (Meters)} & 
\makecell{$\dot{\varphi}$ \\ (Seconds$^{-1}$)} & 
\makecell{$a$ \\ (Meters)} & 
\makecell{$Q_L^2+H_L^2$ \\ (Meters)} \\
\hline
$ 10^{0}$  & $-1.238 \times 10^{-6}$ & $-4.444 \times 10^{3}$  & $1.983 \times 10^{6}$ \\
$ 10^{1}$  & $-1.238 \times 10^{-5}$ & $-4.444 \times 10^{4}$  & $-1.757 \times 10^{7}$ \\
\end{tabular}
\end{ruledtabular}
\end{table}

\section{Conclusions}

Upon analyzing all the characteristics presented in this study, it becomes evident that the $\lambda_5$ solution exhibits the most stable and physically viable behavior, largely due to its asymptotic properties when utilizing a dilatonic scalar field, which satisfies the Null Energy Condition (NEC). In contrast, the $\lambda_{N1}$ solution is only asymptotically flat with a phantom scalar field, which does not fulfill the NEC. However, a detailed examination of the spatial geometry reveals that the type of scalar field does not influence the formation of the wormhole.

Additionally, when examining the behavior of tidal forces for both relativistic and non-relativistic velocities in the $\lambda_5$ solution, the ideal entry angles that prevent disintegration upon passing through the wormhole are near the polar regions $\theta \approx \{ 0,\pi \}$ and, equatorial plane $\theta \approx \pi/2$. It is feasible to traverse the wormhole at a constant $\theta$. Interestingly, there is an absence of tidal forces precisely at the center of the structure ($r=l_1,\theta=0$), for either dilatonic or phantom scalar fields.

Conversely, the $\lambda_{N1}$ solution requires closer attention to tidal force magnitudes for both relativistic and non-relativistic velocities and with either scalar field type, as these forces are bounded. For this solution, it is possible to enter through various regions as long as the peak magnitudes remain within survivable limits for an astronaut. To minimize tidal forces during transit, entry is optimal near the equatorial plane, requiring a variation in $\theta$ to pass through the center of the wormhole. In this case, tidal forces are present in the center of the wormhole.

Plotting the geodesics further confirms that the ring singularity is causally disconnected from these solutions, i.e. the geodesics never reach the ring singularity of the object; this behavior is induced by the geometry of the object itself, as noted in \cite{DelAguila:2021awa}, \cite{DelAguila:2015isj}.

Finally, using the example of WH with solar mass, a pulsar and a super massive black hole and applying relation (\ref{Relacion WH}) in the last chapter, it was determined that the physically most plausible WH is $\lambda_5$, see Table \ref{ValoresL5}. This conclusion arises from the absence of contradictions of the wormhole's parameters and the electromagnetic field, which exhibits a magnitude smaller than that of a magnetar. For the $\lambda_{N1}$ case, it is possible to achieve such a compact object, see Table \ref{ValoresLN1}. However, its throat would be extremely small, approximately 1 meter. Increasing the dimension leads to a contradiction: the sum of the square charges in units of length becomes negative, and the electromagnetic field is practically zero. The field only starts to gain some strength as one moves closer to the WH's center, rendering the existence of a WH with such characteristics unfeasible.

In both cases, the wormhole's rotation is slightly smaller than that of the Sun.

Considering the parameters proposed in this study, which were derived by extending the analysis of black holes to wormholes, among the essential parameters for the characterization of a wormhole are the mass, the electric charge, the magnetic charge, the Schwarzchild radius, the angular momentum per unit mass and the throat size.
\section{Acknowledgements}

LB thanks CONAHCyT-M\'exico for the doctoral grant.
This work was also partially supported by CONACyT M\'exico under grants  A1-S-8742, 304001, 376127, 240512.

\bibliography{Bibliografia}

%apsrev4-2.bst 2019-01-14 (MD) hand-edited version of apsrev4-1.bst
%Control: key (0)
%Control: author (8) initials jnrlst
%Control: editor formatted (1) identically to author
%Control: production of article title (0) allowed
%Control: page (0) single
%Control: year (1) truncated
%Control: production of eprint (0) enabled
\begin{thebibliography}{14}%
\makeatletter
\providecommand \@ifxundefined [1]{%
 \@ifx{#1\undefined}
}%
\providecommand \@ifnum [1]{%
 \ifnum #1\expandafter \@firstoftwo
 \else \expandafter \@secondoftwo
 \fi
}%
\providecommand \@ifx [1]{%
 \ifx #1\expandafter \@firstoftwo
 \else \expandafter \@secondoftwo
 \fi
}%
\providecommand \natexlab [1]{#1}%
\providecommand \enquote  [1]{``#1''}%
\providecommand \bibnamefont  [1]{#1}%
\providecommand \bibfnamefont [1]{#1}%
\providecommand \citenamefont [1]{#1}%
\providecommand \href@noop [0]{\@secondoftwo}%
\providecommand \href [0]{\begingroup \@sanitize@url \@href}%
\providecommand \@href[1]{\@@startlink{#1}\@@href}%
\providecommand \@@href[1]{\endgroup#1\@@endlink}%
\providecommand \@sanitize@url [0]{\catcode `\\12\catcode `\$12\catcode `\&12\catcode `\#12\catcode `\^12\catcode `\_12\catcode `\%12\relax}%
\providecommand \@@startlink[1]{}%
\providecommand \@@endlink[0]{}%
\providecommand \url  [0]{\begingroup\@sanitize@url \@url }%
\providecommand \@url [1]{\endgroup\@href {#1}{\urlprefix }}%
\providecommand \urlprefix  [0]{URL }%
\providecommand \Eprint [0]{\href }%
\providecommand \doibase [0]{https://doi.org/}%
\providecommand \selectlanguage [0]{\@gobble}%
\providecommand \bibinfo  [0]{\@secondoftwo}%
\providecommand \bibfield  [0]{\@secondoftwo}%
\providecommand \translation [1]{[#1]}%
\providecommand \BibitemOpen [0]{}%
\providecommand \bibitemStop [0]{}%
\providecommand \bibitemNoStop [0]{.\EOS\space}%
\providecommand \EOS [0]{\spacefactor3000\relax}%
\providecommand \BibitemShut  [1]{\csname bibitem#1\endcsname}%
\let\auto@bib@innerbib\@empty
%</preamble>
\bibitem [{\citenamefont {Morris}\ and\ \citenamefont {Thorne}(1988)}]{Morris:1988cz}%
  \BibitemOpen
  \bibfield  {author} {\bibinfo {author} {\bibfnamefont {M.~S.}\ \bibnamefont {Morris}}\ and\ \bibinfo {author} {\bibfnamefont {K.~S.}\ \bibnamefont {Thorne}},\ }\bibfield  {title} {\bibinfo {title} {{Wormholes in space-time and their use for interstellar travel: A tool for teaching general relativity}},\ }\href {https://doi.org/10.1119/1.15620} {\bibfield  {journal} {\bibinfo  {journal} {Am. J. Phys.}\ }\textbf {\bibinfo {volume} {56}},\ \bibinfo {pages} {395} (\bibinfo {year} {1988})}\BibitemShut {NoStop}%
\bibitem [{\citenamefont {Gonzalez}\ \emph {et~al.}(2009)\citenamefont {Gonzalez}, \citenamefont {Guzman},\ and\ \citenamefont {Sarbach}}]{Gonzalez:2008wd}%
  \BibitemOpen
  \bibfield  {author} {\bibinfo {author} {\bibfnamefont {J.~A.}\ \bibnamefont {Gonzalez}}, \bibinfo {author} {\bibfnamefont {F.~S.}\ \bibnamefont {Guzman}},\ and\ \bibinfo {author} {\bibfnamefont {O.}~\bibnamefont {Sarbach}},\ }\bibfield  {title} {\bibinfo {title} {{Instability of wormholes supported by a ghost scalar field. I. Linear stability analysis}},\ }\href {https://doi.org/10.1088/0264-9381/26/1/015010} {\bibfield  {journal} {\bibinfo  {journal} {Class. Quant. Grav.}\ }\textbf {\bibinfo {volume} {26}},\ \bibinfo {pages} {015010} (\bibinfo {year} {2009})},\ \Eprint {https://arxiv.org/abs/0806.0608} {arXiv:0806.0608 [gr-qc]} \BibitemShut {NoStop}%
\bibitem [{\citenamefont {Matos}\ and\ \citenamefont {Nunez}(2006)}]{Matos:2005uh}%
  \BibitemOpen
  \bibfield  {author} {\bibinfo {author} {\bibfnamefont {T.}~\bibnamefont {Matos}}\ and\ \bibinfo {author} {\bibfnamefont {D.}~\bibnamefont {Nunez}},\ }\bibfield  {title} {\bibinfo {title} {{Rotating scalar field wormhole}},\ }\href {https://doi.org/10.1088/0264-9381/23/13/012} {\bibfield  {journal} {\bibinfo  {journal} {Class. Quant. Grav.}\ }\textbf {\bibinfo {volume} {23}},\ \bibinfo {pages} {4485} (\bibinfo {year} {2006})},\ \Eprint {https://arxiv.org/abs/gr-qc/0508117} {arXiv:gr-qc/0508117} \BibitemShut {NoStop}%
\bibitem [{\citenamefont {Matos}(2010)}]{Matos:2010pcd}%
  \BibitemOpen
  \bibfield  {author} {\bibinfo {author} {\bibfnamefont {T.}~\bibnamefont {Matos}},\ }\bibfield  {title} {\bibinfo {title} {{Class of Einstein-Maxwell Phantom Fields: Rotating and Magnetised Wormholes}},\ }\href {https://doi.org/10.1007/s10714-010-0976-6} {\bibfield  {journal} {\bibinfo  {journal} {Gen. Rel. Grav.}\ }\textbf {\bibinfo {volume} {42}},\ \bibinfo {pages} {1969} (\bibinfo {year} {2010})},\ \Eprint {https://arxiv.org/abs/0902.4439} {arXiv:0902.4439 [gr-qc]} \BibitemShut {NoStop}%
\bibitem [{\citenamefont {Matos}\ \emph {et~al.}(2000{\natexlab{a}})\citenamefont {Matos}, \citenamefont {Nunez},\ and\ \citenamefont {Rios}}]{Matos:2000za}%
  \BibitemOpen
  \bibfield  {author} {\bibinfo {author} {\bibfnamefont {T.}~\bibnamefont {Matos}}, \bibinfo {author} {\bibfnamefont {D.}~\bibnamefont {Nunez}},\ and\ \bibinfo {author} {\bibfnamefont {M.}~\bibnamefont {Rios}},\ }\bibfield  {title} {\bibinfo {title} {{Class of Einstein-Maxwell dilatons, an ansatz for new families of rotating solutions}},\ }\href {https://doi.org/10.1088/0264-9381/17/18/323} {\bibfield  {journal} {\bibinfo  {journal} {Class. Quant. Grav.}\ }\textbf {\bibinfo {volume} {17}},\ \bibinfo {pages} {3917} (\bibinfo {year} {2000}{\natexlab{a}})},\ \Eprint {https://arxiv.org/abs/gr-qc/0008068} {arXiv:gr-qc/0008068} \BibitemShut {NoStop}%
\bibitem [{\citenamefont {Matos}\ \emph {et~al.}(2016)\citenamefont {Matos}, \citenamefont {Urena-Lopez},\ and\ \citenamefont {Miranda}}]{Matos:2012gj}%
  \BibitemOpen
  \bibfield  {author} {\bibinfo {author} {\bibfnamefont {T.}~\bibnamefont {Matos}}, \bibinfo {author} {\bibfnamefont {L.~A.}\ \bibnamefont {Urena-Lopez}},\ and\ \bibinfo {author} {\bibfnamefont {G.}~\bibnamefont {Miranda}},\ }\bibfield  {title} {\bibinfo {title} {{Wormhole Cosmic Censorship}},\ }\href {https://doi.org/10.1007/s10714-016-2040-7} {\bibfield  {journal} {\bibinfo  {journal} {Gen. Rel. Grav.}\ }\textbf {\bibinfo {volume} {48}},\ \bibinfo {pages} {61} (\bibinfo {year} {2016})},\ \Eprint {https://arxiv.org/abs/1203.4801} {arXiv:1203.4801 [gr-qc]} \BibitemShut {NoStop}%
\bibitem [{\citenamefont {Del~\'Aguila}\ and\ \citenamefont {Matos}(2021)}]{DelAguila:2021awa}%
  \BibitemOpen
  \bibfield  {author} {\bibinfo {author} {\bibfnamefont {J.~C.}\ \bibnamefont {Del~\'Aguila}}\ and\ \bibinfo {author} {\bibfnamefont {T.}~\bibnamefont {Matos}},\ }\bibfield  {title} {\bibinfo {title} {{Gravitational perturbations in the Newman-Penrose formalism: Applications to wormholes}},\ }\href {https://doi.org/10.1103/PhysRevD.103.084033} {\bibfield  {journal} {\bibinfo  {journal} {Phys. Rev. D}\ }\textbf {\bibinfo {volume} {103}},\ \bibinfo {pages} {084033} (\bibinfo {year} {2021})},\ \Eprint {https://arxiv.org/abs/2102.09160} {arXiv:2102.09160 [gr-qc]} \BibitemShut {NoStop}%
\bibitem [{Note1()}]{Note1}%
  \BibitemOpen
  \bibinfo {note} {Taken from the article \cite {Matos:2010pcd}.}\BibitemShut {Stop}%
\bibitem [{\citenamefont {Hans}\ \emph {et~al.}(2009)\citenamefont {Hans}, \citenamefont {Dietrich}, \citenamefont {MacCallum}, \citenamefont {Hoenselaers},\ and\ \citenamefont {Herlt}}]{Cita:ExactSolutions}%
  \BibitemOpen
  \bibfield  {author} {\bibinfo {author} {\bibfnamefont {S.}~\bibnamefont {Hans}}, \bibinfo {author} {\bibfnamefont {K.}~\bibnamefont {Dietrich}}, \bibinfo {author} {\bibfnamefont {M.}~\bibnamefont {MacCallum}}, \bibinfo {author} {\bibfnamefont {C.}~\bibnamefont {Hoenselaers}},\ and\ \bibinfo {author} {\bibfnamefont {E.}~\bibnamefont {Herlt}},\ }\href {https://doi.org/https://doi.org/10.1017/CBO9780511535185} {\emph {\bibinfo {title} {Exact Solutions of Einstein's Field Equations}}},\ \bibinfo {edition} {2nd}\ ed.\ (\bibinfo  {publisher} {Cambridge University Press},\ \bibinfo {year} {2009})\BibitemShut {NoStop}%
\bibitem [{Note2()}]{Note2}%
  \BibitemOpen
  \bibinfo {note} {For a comprehensive derivation, refer to \cite {Matos:2000ai}.}\BibitemShut {Stop}%
\bibitem [{Note3()}]{Note3}%
  \BibitemOpen
  \bibinfo {note} {The Hermite method is a high-order interpolation method suitable for smooth solutions, automatically selected in NDSolve when the equations and integration domain meet specific smoothness and stability conditions.}\BibitemShut {Stop}%
\bibitem [{Note4()}]{Note4}%
  \BibitemOpen
  \bibinfo {note} {There exists a transcendental condition, $L<2l_1$, i.e., the throat is situated within the area defined by the Schwarzschild radius.}\BibitemShut {Stop}%
\bibitem [{\citenamefont {Del~\'Aguila}\ \emph {et~al.}(2019)\citenamefont {Del~\'Aguila}, \citenamefont {Matos},\ and\ \citenamefont {Miranda}}]{DelAguila:2015isj}%
  \BibitemOpen
  \bibfield  {author} {\bibinfo {author} {\bibfnamefont {J.~C.}\ \bibnamefont {Del~\'Aguila}}, \bibinfo {author} {\bibfnamefont {T.}~\bibnamefont {Matos}},\ and\ \bibinfo {author} {\bibfnamefont {G.}~\bibnamefont {Miranda}},\ }\bibfield  {title} {\bibinfo {title} {{Exact Rotating Magnetic Traversable Wormholes satisfying the Energy Conditions}},\ }\href {https://doi.org/10.1103/PhysRevD.99.124045} {\bibfield  {journal} {\bibinfo  {journal} {Phys. Rev. D}\ }\textbf {\bibinfo {volume} {99}},\ \bibinfo {pages} {124045} (\bibinfo {year} {2019})},\ \Eprint {https://arxiv.org/abs/1507.02348} {arXiv:1507.02348 [gr-qc]} \BibitemShut {NoStop}%
\bibitem [{\citenamefont {Matos}\ \emph {et~al.}(2000{\natexlab{b}})\citenamefont {Matos}, \citenamefont {Nunez}, \citenamefont {Estevez},\ and\ \citenamefont {Rios}}]{Matos:2000ai}%
  \BibitemOpen
  \bibfield  {author} {\bibinfo {author} {\bibfnamefont {T.}~\bibnamefont {Matos}}, \bibinfo {author} {\bibfnamefont {D.}~\bibnamefont {Nunez}}, \bibinfo {author} {\bibfnamefont {G.}~\bibnamefont {Estevez}},\ and\ \bibinfo {author} {\bibfnamefont {M.}~\bibnamefont {Rios}},\ }\bibfield  {title} {\bibinfo {title} {{Rotating 5-D Kaluza-Klein space-times from invariant transformations}},\ }\href {https://doi.org/10.1023/A:1001982001694} {\bibfield  {journal} {\bibinfo  {journal} {Gen. Rel. Grav.}\ }\textbf {\bibinfo {volume} {32}},\ \bibinfo {pages} {1499} (\bibinfo {year} {2000}{\natexlab{b}})},\ \Eprint {https://arxiv.org/abs/gr-qc/0001039} {arXiv:gr-qc/0001039} \BibitemShut {NoStop}%
\end{thebibliography}%


%apsrev4-2.bst 2019-01-14 (MD) hand-edited version of apsrev4-1.bst
%Control: key (0)
%Control: author (8) initials jnrlst
%Control: editor formatted (1) identically to author
%Control: production of article title (0) allowed
%Control: page (0) single
%Control: year (1) truncated
%Control: production of eprint (0) enabled
%
\end{document}